\documentclass[a4paper,fleqn,usenatbib]{mnras}
\usepackage{graphicx}	
\usepackage{amsmath}	
\usepackage{amssymb}	

\def\msun{{\rm M_{\odot}}}

\def\be{\begin{equation}}
\def\ee{\end{equation}}

\def\del#1{{}}

\newcommand\mj{{\,{\rm M}_{\rm J}}}

\title[Extreme evaporation of planets in FUORs]{Extreme evaporation of planets in hot thermally unstable protoplanetary discs: the case of FU Ori}

\author[Nayakshin et al.]{Sergei Nayakshin$^{1}$, James E. Owen$^{2}$, Vardan Elbakyan$^{1}$\\
$^{1}$School of Physics and Astronomy, University of
  Leicester, Leicester LE1 7RH, UK.\\
  $^{2}$Astrophysics Group, Department of Physics, Imperial College London, Prince Consort Rd, London SW7 2AZ, UK
}

\date{Accepted XXX. Received YYY; in original form ZZZ}

\pubyear{2022}

\begin{document}
\label{firstpage}
\pagerange{\pageref{firstpage}--\pageref{lastpage}}
\maketitle

\begin{abstract}
Disc accretion rate onto low mass protostar FU Ori suddenly increased hundreds of times 85 years ago and remains elevated to this day. We show that the sum of historic and recent observations challenges existing FU Ori models. We build a theory of a new process, Extreme Evaporation (EE) of young gas giant planets in discs with midplane temperatures of $\gtrsim 30,000$~K. Such temperatures are reached in the inner $0.1$ AU during thermal instability bursts. In our 1D time-dependent code the disc and an embedded planet interact through gravity, heat, and mass exchange. We use disc viscosity constrained by simulations and observations of dwarf novae instabilities, and we constrain planet properties with a stellar evolution code. We show that dusty gas giants born in the outer self-gravitating disc reach the innermost disc in a $\sim O(10^4)$ years with radius of $\sim 10 R_{\rm J}$. We show that their EE rates are $\gtrsim 10^{-5} \msun$~yr$^{-1}$; if this exceeds the background disc accretion activity then the system enters a planet-sourced mode. Like a stellar secondary in mass-transferring binaries, the planet becomes the dominant source of matter for the star, albeit for $\sim O(100)$ years. We find that a $\sim 6$ Jupiter mass planet evaporating in a disc fed at a time-averaged rate of $\sim 10^{-6} \msun$~year$^{-1}$ appears to explain all that we currently know about FU Ori accretion outburst. More massive planets and/or planets in older less massive discs do not experience EE process. Future FUOR modelling may constrain planet internal structure and evolution of the earliest discs.

\end{abstract}


\section{Introduction}

The prototypical star FU Orionis is a member of the FUOR class \citep{Herbig89}, young accreting stars that undergo extreme, short increases in brightness by as much as 4-6 magnitudes followed by slow decay on timescales of tens to hundreds of years \citep[for reviews see][]{AudardEtal14,Fischer-PPVII}. \cite{HK96} argued that most low mass stars go through the FUOR stage, experiencing bursts of accretion rate $\dot M_*$ as high as $10^{-4} \msun$~yr$^{-1}$. While spectral characteristics of FUORs clearly implicate disc accretion as outbursts origin \citep[e.g., see][]{Zhu09-FUOri-obs}, the exact physics of the disc that leads to episodic accretion remains a puzzle.

A number of authors \citep[e.g.,][]{HartmannK85-FUORs,Clarke-90-FUOR}  proposed a thermal instability (TI) scenario for FUORs \citep[by analogy with dwarf novae outbursts][]{Meyer84-CVs,Smak84-dwarf-novae-review}. \cite{Bell94} built detailed TI models for FUORs, showing that for a range of  $\dot M_{\rm feed}$, the mass accretion rates at which the inner disc is fed from outside, the inner disc must show a cyclic behaviour, switching between a stable ``cold"  branch and a stable ``hot''  one. The former is characterised by disc temperatures $T\lesssim (2-3) \times 10^3$~K, when Hydrogen in the disc is neutral. The hot branch has $T$ closer to $10^4$~K, with Hydrogen in the disc fully ionised. On the cold branch, the inner disc viscosity is low, and accretion rate, $\dot M = \dot M_{\rm c}$, is lower than $\dot M_{\rm feed}$. This forces the gas to pile up in the disc at a distance of $R\lesssim 0.1 AU$ from the star. When enough matter accumulates, the disc heats up for Hydrogen to become ionised. On the hot branch, the viscosity is high, and the accretion rate $\dot M_{\rm h}$ greatly exceeds  $\dot M_{\rm feed}$. The inner disc is thus drained onto the star in an outburst, the disc  returns onto the cold branch, and the cycle repeats.



Further work \citep[e.g.,][]{LodatoClarke04} unfortunately uncovered significant challenges for the classic TI scenario \citep[for a review see][his \S 6 and \S 6.1.2]{Armitage15-review}. To match the observed long outburst durations ($\gtrsim 100$~years), very low values of the viscosity parameter $\alpha$ are required, from $\sim 10^{-4}$ to $10^{-3}$. MRI simulations of the inner completely ionised disc \citep{Davis-10-MRI,Simon-12-MRI,Hirose15,Coleman16-TI} and observational evidence \citep{King07-alpha,Hameury-20-review} suggest much higher values, $\alpha \sim 10^{-2} - 10^{-1}$. The extent of the active disc region in these models, $R_{\rm ac}\lesssim 0.1$~AU, appeared too small compared with FU Ori SED modelling and interferometric observations \citep[which found $R_{\rm ac}$ between 0.5 and 1 AU, see][]{ZhuEtal07,Zhu08-FUOri,Zhu09-FUOri-obs,Eisner11-FUOR,Labdon21-FUOR}. 


\cite{Clarke96-FUOR,LodatoClarke04} showed that a  massive gas giant planet migrating inward through the disc affects it strongly, potentially triggering thermal instability further out in the disc than in the classic \cite{Bell94} models. However, the main challenge of this planet-TI model is that massive ($\sim 10\mj$) gas giants are observed to be excessively rare. Using the observed mass function of gas giants \citep{CummingEtal08}, one may estimate the rate of such planets to be of order   $\sim 0.1$\% for FGK stars given the $\sim 1$\% frequency of all hot jupiters \citep{SanterneEtal15}.

A clump migration scenario for FUORs was pioneered by \cite{VB05,VB06,VB10} who showed that massive circumstellar discs fragment due to gravitational instability (GI), producing numerous self-gravitating gas clumps. If these clumps migrate all the way to the star then intense FUOR-like accretion bursts occur. These ideas were re-invented in the field of planet formation when several authors showed that clump tidal disruptions might yield not only FUOR-like episodes of high mass accretion rate onto the stars but also leave behind sub-Jovian mass planets or solid cores \citep[e.g.,][]{BoleyEtal10,Nayakshin10c,ChaNayakshin11a}. In relation to FUORs specifically, planet disruptions must occur in the very inner disc $R \sim (0.1-0.5)$~AU or else the outburst durations and rise times are too long compared to observations. \cite{NayakshinLodato12}  found that planets need to be  radially extended (several tens of Jupiter radii, $R_J$) to provide better agreement with observations. 
However, classical FUOR outbursts are human lifetime long, and generally surprisingly steady \citep[e.g., fig. 4 in][]{Fischer-PPVII}. Tidal disruption of planets produces short ($\sim$ years long) spiky outbursts unless the planet is able to carve a deep gap in the disc {\em during} the outburst. 
This would only be possible is $\alpha \sim 10^{-3}$ in the inner hot disc.


An MRI activation scenario is another model for FUORs \citep{ArmitageEtal01,Bae13-MRI-1D,ZhuEtal09,Bae14-MRI-2D}. \cite{Gammie96} showed that for sufficiently small disc accretion rates, regions between a fraction of an AU to a few AU may be too cold (central disc temperature $T_{\rm c} \lesssim 10^3$~K) to sustain a sufficient degree of ionisation to support the MRI turbulence \citep{Balbus91}. Only the upper layers of such discs are ionised and transport the matter towards the star at rates $\dot M_{\rm q} \sim 10^{-8}\msun$~year$^{-1}$. If $\dot M_{\rm feed} \gg \dot M_{\rm q}$, the matter piles up in the disc at a few AU distance from the star. Eventually, a mass reservoir with mass $\sim 0.1\msun$ accumulates at these distances. Residual hydrodynamic turbulence and/or the disc's self-gravity then heats the disc enough to force alkaline metal ionisation. As the dead zone viscosity increases strongly, the disc inward of $\sim 3$~AU becomes active. A fraction of the dead zone mass is dumped onto the star; an accretion outburst results. 

Recently \cite{2022Lykou} observed FU Ori via MATISSE/VLT interferometry in {\em L, M, N} bands and also obtained contemporaneous photometry in the number of optical and near-infra-red bands. Their results indicate that the actively accreting disc is $R_{\rm ac} =0.3$~AU, significantly smaller than previous estimates (see \S \ref{sec:R_disc} for more discussion of this result). We show in \S \ref{sec:R_disc_MRI} that this is an order of magnitude too small for the classical MRI-activation scenario, requiring that $\alpha\lesssim 10^{-4}$ in the hot ionised region of the disc. 


Further, photometric variability of FU Ori on timescales of days to $\sim 14$ days led to suggestions of a massive planet orbiting the star  \citep[e.g.,][]{PowellEtal12}. \cite{Siwak21-FUOri-QPOs} used the MOST satellite data, avoiding the problems of unevenly sampled weather-depending ground-based data, firming up the $\sim (10-12)$ days quasi-periodic variability of FU Ori with amplitude of up to 0.07 magnitudes. For the stellar mass of $M_* = 0.6\msun$ \citep{Perez20-FUOri}, this period corresponds to a circular orbit at $R \approx 0.08$~AU.


In this paper take a closer look at the planet-TI scenario. Here we consider not only gravitational \citep{LodatoClarke04} and mass-deposition \citep{NayakshinLodato12} coupling between the disc and the planet but also thermal effects of the disc onto the planet. \cite{VazanHelled12} demonstrated that very young pre-collapse gas giant planets (for terminology see \S \ref{sec:planet_origin}) could be unbound by overheating in  the thermal bath of the disc at distances $\sim$ 5-10 AU from the star. Post-collapse planets are orders of magnitude denser and hotter \citep{Bodenheimer74,GraboskeEtal75}, yet we show that they too are vulnerable to the thermal bath effects in the inner disc during TI bursts when they are exposed to temperatures $T \gtrsim 3\times 10^4$~K. 

In \S 3 \& 4 we argue that recent FU Ori observations challenge the MRI activation or any other scenario in which matter feeding FU Ori comes from regions larger than a fraction of an AU; a steady-state source placed  inside the observed active disc region is required. In \S 5 we provide a physical model describing such a mass source. In \S 6 we present numerical experiments of increasing complexity to show how the planet and the disc interact. We then build in \S 7 a detailed model for the FU Ori outburst, constraining the disc and the planet properties tightly. A brief discussion of the model in the context of observations is given in \S 8.



\section{Numerical method}\label{sec:numerical_method}

We build on a time-dependent 1D viscously evolving disc model with an optionally embedded planet in it \citep{NayakshinLodato12,Nayakshin22-ALMA-CA},
\begin{equation}
\begin{split}
    \frac{\partial\Sigma}{\partial t} = \frac{3}{R} \frac{\partial}{\partial R} \left[ R^{1/2} \frac{\partial}{\partial R} \left(R^{1/2}\nu \Sigma\right) \right] - \frac{1}{R} \frac{\partial}{\partial R} \left(2\Omega^{-1} \lambda \Sigma\right) + \\ 
    \frac{\dot{M}_{\rm p}}{2\pi R} D(R - a)\;,
\end{split}
\label{dSigma_dt}
\end{equation}
where $\Sigma$ is the disc surface density, $\Omega$ is the angular velocity, $\nu=\alpha c_{\rm{s}} H$ is the \cite{Shakura73} kinematic viscosity, $c_{\rm{s}}$ is the midplane sound speed, and $H$ is the disc vertical scale height. The last two terms in eq. \ref{dSigma_dt} describe the angular momentum exchange between the disc and the planet via tidal torques, and the planet mass loss (that is deposited into the surrounding disc), respectively. The function $D(R-a)$ is a narrow Gaussian; when integrated over the disc area, the last term yields the planetary mass loss rate, $\dot M_{\rm p}$, discussed in  \S \ref{sec:EE}. The expressions for the planet-disc tidal torques are given in \cite{Nayakshin22-ALMA-CA}. These act on the planet (forcing it to migrate) and on the disc (modifying the disc surface density at high planet masses). A smooth transition between the type I and type II migration regimes is applied.

For quantitatively accurate results, energy equilibrium in TI-unstable inner discs needs to be solved through the vertical energy heating-cooling balance equation \citep[e.g.,][]{Bell94, Hameury98-alpha, Lasota08-S-curve}. These detailed models depend on the treatment of convection and assumptions about how the heating is distributed vertically. \cite{Hirose14-Convection,Hirose15,Coleman16-TI} perform 3D radiation-magneto-hydrodynamic simulations of discs in shearing boxes. They find that convection contributes significantly to the MRI turbulence on the hot branch and explain from first principles why discs in dwarf novae require $\alpha_{\rm cold} \sim 0.01$ on the cold branch but $\alpha_{\rm hot} \sim 0.1$ on the hot one \citep[e.g.,][]{Lasota01-Review,Hameury-20-review}. In this paper, we use the one-zone approximation to the vertical heating-cooling balance of thin discs \citep[e.g.,][]{LodatoClarke04,Zhu10-DZ-MRI-1D,Bae13-MRI-1D,VB15,Kadam20-DZ-MRI}. The disc midplane temperature is solved for via:
\begin{equation}
    \frac{d T_{\rm d}}{dt} = - \frac{T_{\rm d} - T_{\rm eq}}{t_{\rm therm}} - \frac{1}{R} \frac{\partial}{\partial R}\Big[ T_{\rm d} R v_R  \Big]\;,
    \label{energy_equation}
\end{equation}
where $t_{\rm therm} = (\alpha \Omega)^{-1}$ is the local disc thermal time scale, and $T_{\rm eq}$ is the equilibrium disc temperature found by balancing the radiative cooling rate with the local viscous heating rate. The last term in eq. \ref{energy_equation} describes radial advection of heat in the disc, with $v_R$ the radial gas velocity in the disc.


\section{Why is the active disc so small?}\label{sec:R_disc}


\cite{2022Lykou} presented MATISSE/VLT interferometry in {\em L, M, N} bands and  contemporaneous photometry of FU Ori in the number of optical and near-infra-red bands. Their superior spatial resolution and detailed radiative transfer modelling show an unexpectedly small radius for the hot actively accreting disc, $R_{\rm ac} \sim 0.3$~AU. The source is only marginally resolved at the longest baselines at 3.5 microns ({\em L} band), yielding the emitting region at 3.5 $\mu$m of $\lesssim 0.25$~AU. According to Wien's displacement law, the {\em L} band emission peaks at temperature $\sim 800$ K. A steady-state self-luminous accretion disc would have this temperature at $R\approx 0.6$~AU at accretion rate $\dot M = 3.8\times 10^{-5}\msun$~year$^{-1}$ \citep{Perez20-FUOri} (in \S \ref{sec:Radial_flux_distribution} the outer edge of the disc in L band in this scenario is quantified to even larger value, $R_{\rm ac} \sim 0.9$~AU). Second, \cite{2022Lykou} built a Monte Carlo radiative transfer model, breaking the disc into an actively accreting inner disc and a passive outer disc, and finding $R_{\rm ac}\approx 0.3$~AU. 



\subsection{MRI activation scenario}\label{sec:R_disc_MRI}

In quiescence, a mass reservoir exists in the disc in the ``dead zone", where the viscosity is negligible \citep{ArmitageEtal01}. At the beginning of the outburst, the gas temperature exceeds a critical temperature, and the dead zone is ionised.  One testable prediction of the model is $R_{\rm ac}$, the size of the region that participates in the outburst. The  the 1D \citep[e.g.,][]{ArmitageEtal01,Zhu10-DZ-MRI-1D,Bae13-MRI-1D}, 2D \citep{Bae14-MRI-2D,Kadam20-DZ-MRI} and 3D \citep{Zhu20-FUOR} simulations of the MRI activation scenario all show that $R_{\rm ac}$ is a few AU. Here we make a simple yet robust argument that $R_{\rm ac}$ has to be that large in this scenario.  Consider the hot active disc as a steady-state disc extending from the star to radius $R_{\rm ac}$. Since the luminosity of FU Ori varied little over $\sim 85$ years, the mass of gas in the active zone is $M_{\rm ac} \gtrsim \dot M_* \times 85\, {\rm years} \geq 2\times 10^{-3}\msun$. The disc surface density at $R_{\rm ac}$ is 
\begin{equation}
    \Sigma \gtrsim \frac{M_{\rm ac}}{\pi R_{\rm ac}^2} = 5.66\times 10^3 R_1^{-2}  \; {\rm g~cm}^{-2}\;,
    \label{Sigma_ac}
\end{equation}
where $R_1 = R_{\rm ac}/(1$~AU). The disc optical depth is $\tau = \kappa_{\rm R} \Sigma/2 \gg 1$, where $\kappa_{\rm R}$ is Rosseland opacity coefficient. The balance of viscous heating and radiative cooling gives
\begin{equation}
    \frac{3 G M_*\dot M_*}{8\pi R_{\rm ac}^3} = \frac{\sigma_{\rm B} T_{\rm d}^4}{\tau} =
    \sigma_{\rm B}T_{\rm d}^4 \frac{2 \pi R_{\rm ac}^2}{\kappa_{\rm R} M_{\rm ac}}\;,
    \label{Thermal_balance}
\end{equation}
where $\sigma_{\rm B}$ is the Stefan-Boltzmann constant, and $T_{\rm d}$ is the disc midplane temperature. Assuming that the MRI turbulence is revived at critical temperature $T_{\rm ac}\sim 10^3$~K, we solve for the radius where $T_{\rm d}$ first falls below $T_{\rm ac}$:
\begin{equation}
    R_{\rm ac} =  \left(\frac{3 G M_*\dot M_*}{16\pi^2\sigma_{\rm B} T_{\rm ac}^4} \kappa_{\rm R} M_{\rm ac}\right)^{1/5} = 
    2.8 \,{\rm AU} \; \kappa_{\rm R}^{1/5}T_3^{-4/5}\;,
    \label{Racc0}
\end{equation}
where $T_3 = T_{\rm ac}/(10^{3}$~K). The result does not depend on $\alpha$.

Non-ideal magnetohydrodynamics simulations of protoplanetary discs \citep[e.g.,][]{BaiStone13,Lesur21-MHD-winds} indicate that the disc angular momentum transfer can be dominated by MHD winds \citep{BlandfordPayne82}. In this case, energy generation may occur high up in the disc atmosphere, not in the midplane. Due to this, and also due to energy losses for launching winds, such discs may be much cooler \citep[e.g.,][]{Suzuki16-MHD-winds,Mori-19-MHD-winds} than the standard turbulent viscosity ones.  Could these effects be sufficiently strong to let the disc in FU Ori have midplane temperature $T_{\rm d} = T_{\rm ac} \sim 10^3$~K at the observed active disc edge, $R_{\rm ac}\approx 0.3$~AU, rather than $\sim 3$~AU? To evaluate this idea, let $f_{\rm visc}$ be the fraction of the local accretion flow energy liberation rate dissipated due to MRI turbulence in the disc midplane. The lower $f_{\rm visc}$, the cooler the disc in the midplane, and so we can constrain the maximum $f_{\rm visc}$ that would satisfy $T_{\rm d}\lesssim T_{\rm ac}$ at 0.3 AU. In this scenario eq. \ref{Thermal_balance} can be re-written:
 \begin{equation}
   f_{\rm visc} \frac{3 G M_*\dot M_*}{8\pi R_{\rm ac}^3} =  \sigma_{\rm B}T_{\rm d}^4 \frac{2 \pi R_{\rm ac}^2}{\kappa_{\rm R} M_{\rm ac}}\;.
    \label{Thermal_balance-1}
\end{equation}
With $T_{\rm d} = T_{\rm ac}$, we find
\begin{equation}
    f_{\rm visc} \lesssim 4\times 10^{-5}  \left(\frac{R_{\rm ac}}{\rm 0.3 AU}\right)^{5}
    T_3^4 \kappa_{\rm R}^{-1}\;.
    \label{f_visc}
\end{equation}
This is vanishingly small and unlikely for an MRI-active hot inner disc. Indeed, in this case, $1-f_{\rm visc} \sim 1$, so $\approx 100$\% of the accretion flow would be due to a magnetised disc wind. If we parameterise the efficiency of angular momentum extraction via an $\alpha_{\rm dw}$ prescription \citep{Tabone22-general}, then eq. \ref{f_visc} implies that the turbulent viscosity $\alpha \sim f_{\rm visc} \alpha_{\rm dw}$ is exceedingly small. For example, at $\alpha_{\rm dw} = 0.1$ we have $\alpha \lesssim 4\times 10^{-6}$, whereas simulations of ionised MRI-active discs show $\alpha\sim 0.1$ \citep[e.g.,][]{Hirose14-Convection,Hirose15,Zhu20-FUOR}.


\subsection{A mass reservoir close to the star}\label{sec:inside_reservoir}

Let the burst be powered by a reservoir of mass located at  $R_{\rm res} \ll R_{\rm ac}$, and that prior to the burst, the disc mass  in the inner regions was very low compared to the reservoir mass. The outer edge of the {\em active} disc, in this case, is where the disc manages to spread  viscously {\em in the outward direction} by a time $t$ (here $t=0$ is defined as the beginning of FU Ori outburst). By order of magnitude, this is where  the viscous time $t_{\rm visc} = 1/(\alpha h^2 \Omega) \approx t$, with $h= H/R$. Numerically,
\begin{equation}
    R_{\rm ac} \approx \left[ t \times \alpha  h^2 (G M_*)^{1/2} 
    \right]^{2/3} = 0.25 \text{AU } \left(\frac{\alpha}{ 0.03} \frac{t}{85 \hbox{ yr}} h_{-1}^2 \right)^{2/3}
    \label{R_ac_planet0}
\end{equation}
where $h_{-1}= h/0.1$.
This estimate is reasonably close to the observed value $R_{\rm ac}\approx 0.3$~AU for a realistic value of $\alpha$.

\subsubsection{An instantaneous mass release in the inner disc}\label{sec:prompt_td}



\begin{figure}
\includegraphics[width=1\columnwidth]{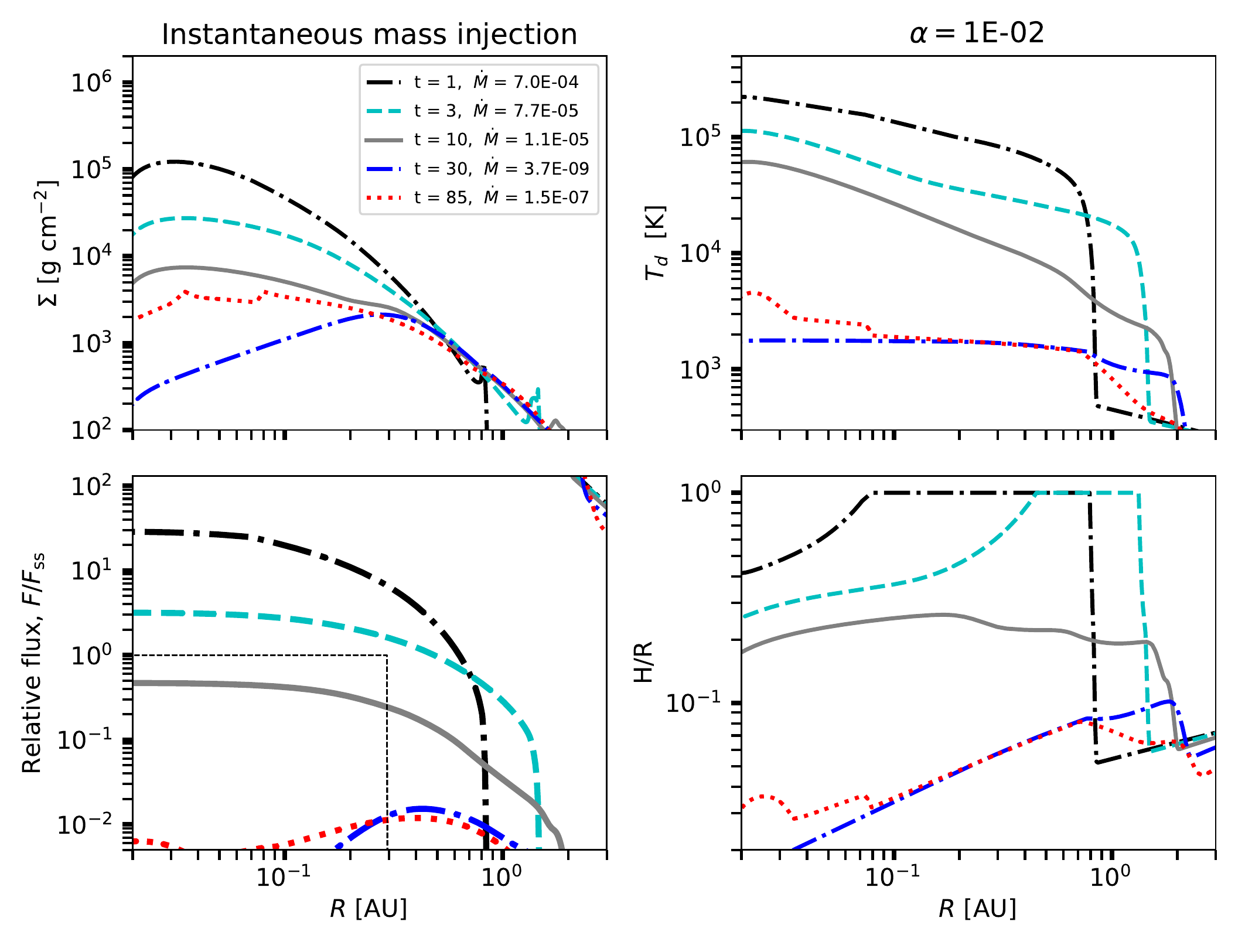}
\caption{An instantaneous injection of $M_{\rm p} = 3\mj$ into the inner disc (see \S \ref{sec:prompt_td}).
Shown are disc surface density $\Sigma$, midplane temperature, radiation flux normalised to the steady-state flux, and the disc aspect ratio, $H/R$. }
\label{fig:Disc_details_TD}
\end{figure}

A sudden tidal disruption of a gas giant planet may result in a sudden injection of matter into the disc. Here we add 
$3\mj$ of gas instantaneously into the disc in a narrow ring centred on radius $R=0.08$~AU to comply with the QPO variability seen by \cite{Siwak21-FUOri-QPOs}. 
Fig. \ref{fig:Disc_details_TD} shows several snapshots of the radial profiles for the disc surface density $\Sigma$, disc midplane temperature, and the aspect ratio $H/R$. This model is computed for \cite{Shakura73} viscosity parameter $\alpha =0.01$. We also show the profile of the relative radiative cooling flux of the disc, defined as the ratio ${\cal F} = F/F_{\rm ss}$, where the steady-state flux $F_{\rm ss}$ is
\begin{equation}
    F_{\rm ss} =  \frac{3 G M_*\dot M_*}{8\pi R^3} j(R)\;,
    \label{F_ss}
\end{equation}
with $j(R) = 1- (R_*/R)^{1/2}$ and stellar accretion rate $\dot M_* = 2\times 10^{-5}\msun$~year$^{-1}$. ${\cal F}$ provides an important insight: if the model disc were in a steady state, and the mass accretion rate as inferred by \cite{2022Lykou}, then ${\cal F} = 1$ up to the outer edge of the active disc where the flux rolls over to pre-outburst values. The black dotted box in the bottom left panel of Fig. \ref{fig:Disc_details_TD}  is the observationally desirable outline of ${\cal F}$.

Fig. \ref{fig:Disc_details_TD} shows that the disc evolves extremely rapidly. While some of the matter accretes onto the star at very high rates, other material spreads viscously outward. The inner disc is as hot as $\sim 10^5$~K, bringing $H/R$ to values formally exceeding unity (we cap $H/R$ at unity to prevent this unphysical situation). However, the disc cools very quickly and, within tens of years, drops from the hot stable state into the cold stable one where the disc temperature is $\sim 2000$~K. This rapid $\dot M_*$ evolution is unlike the observed very slowly declining light curve of FU Ori. Also, the disc flux profile only briefly looks somewhat close to the desired black dotted curve. 

\begin{figure}
\includegraphics[width=1\columnwidth]{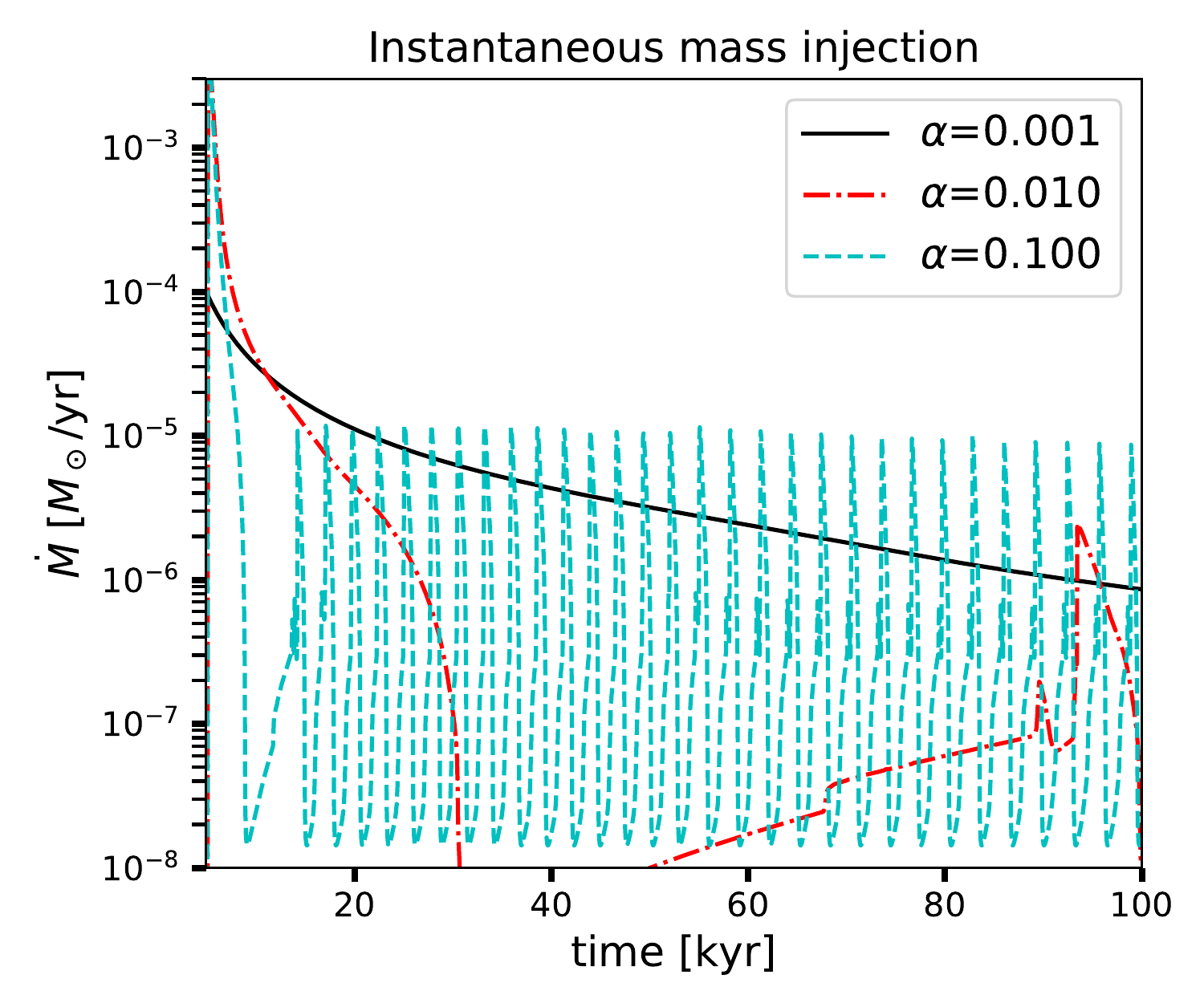}
\caption{Stellar accretion rate for three instantaneous mass injection experiments described in \S \ref{sec:prompt_td}.}
\label{fig:History_TD}
\end{figure}

Fig. \ref{fig:History_TD} shows $\dot M_*$ for this toy experiment and two analogous calculations but for $\alpha = 0.001$ and $\alpha = 0.1$. None of the experiments yields an $\dot M_*$  similar to the observed one. The best match is obtained for an realistically low $\alpha = 0.001$, for which the relative disc flux ${\cal F}$, not present here for brevity, is also not consistent with the observations.  For $\alpha =0.01$, the inner disc drops into quiescence already after 30 years of the burst, although it gets revived somewhat at later times due to TI. The disc evolves even more rapidly for $\alpha = 0.1$ and becomes TI-unstable, very unlike FU Ori. 

\subsubsection{An ad hoc steady state mass source}\label{sec:ad_hoc_steady}

\begin{figure*}
\includegraphics[width=0.9\textwidth]{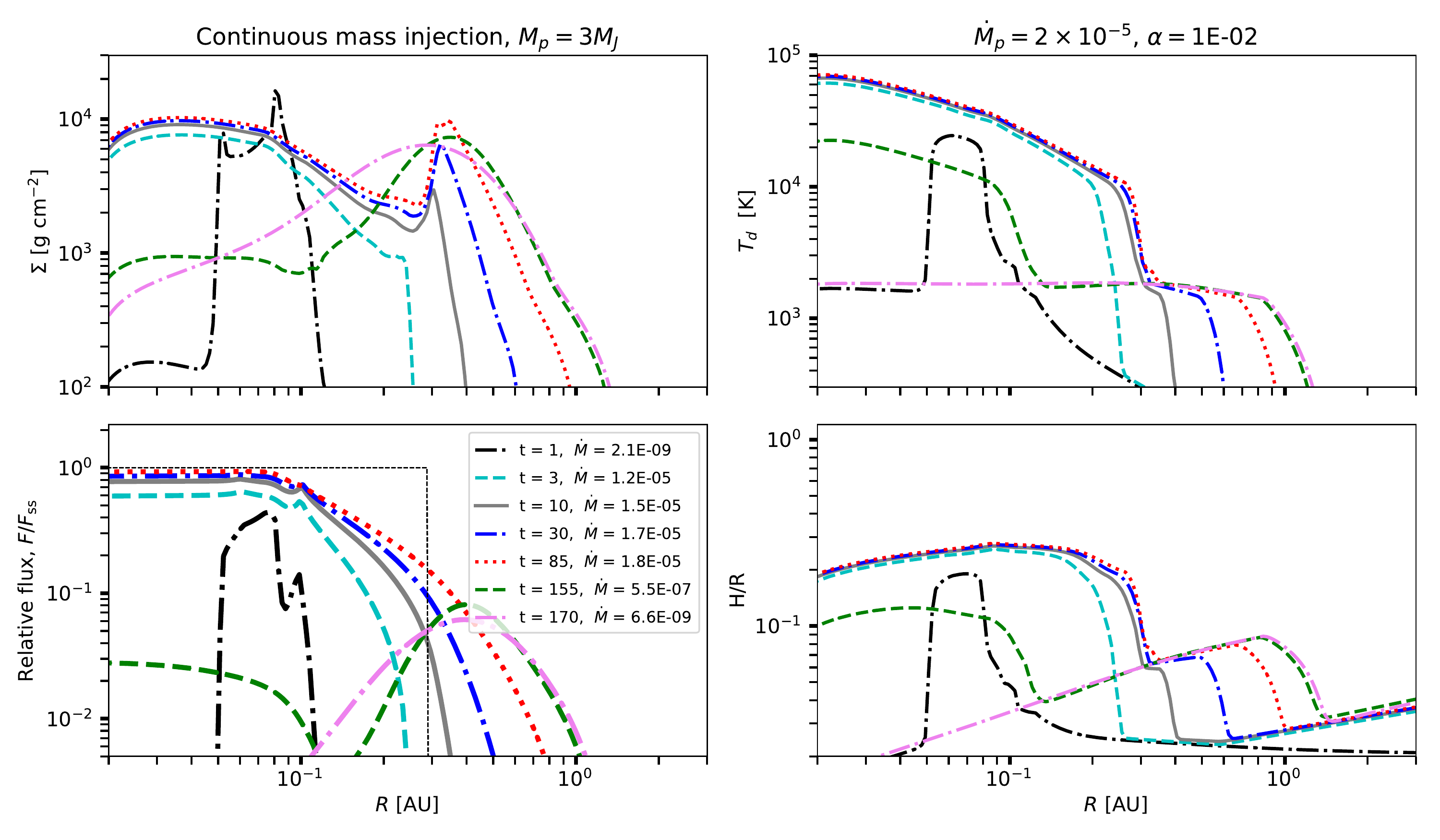}
\caption{Same as Fig. \ref{fig:Disc_details_TD}  but now the planet lose mass at a constant rate of $\dot M_{\rm p} = 2\times 10^{-5} \msun$~year$^{-1}$. The relative flux ${\cal F}$ is close to the desired shape for $\sim 90$\% of the outburst duration.}
\label{fig:Disc_FixedMdot_alpha1m2}
\end{figure*}


Here we experiment with a planet orbiting FU Ori at $R=0.08$~AU which is losing mass at a fixed rate $\dot M_{\rm p} = 2\times 10^{-5}\msun$ yr$^{-1}$. The initial mass of the planet is $M_{\rm p} = 3\mj$. Fig. \ref{fig:Disc_FixedMdot_alpha1m2} shows the disc profiles for a selection of times in the same format as in Fig. \ref{fig:Disc_details_TD}. Encouragingly, the relative flux ${\cal F}$ in Fig. \ref{fig:Disc_FixedMdot_alpha1m2} conforms qualitatively well to the desired step-like shape in a broad time interval, from $t\approx 10$ years to $t=85$ years, and in fact, until the planet runs out of mass at $t\approx 140$ years. Further, we repeated this steady-state planet mass loss experiment with $\alpha=0.1$ and $\alpha = 0.001$. We found that, surprisingly, the size of the active disc in this scenario depends very little on $\alpha$.  \cite{Zhu08-FUOri} concluded from radiative transfer SED modelling that the accretion rate in the passive disc  must drop by at least a factor of 4 compared to that in the active one. If we use the same definition for the size of the active disc here, that is, ${\cal F} \geq 1/4$ in the active disc, then $R_{\rm ac} = $ 0.2, 0.22, 0.32~AU for $\alpha = 10^{-3}$, $10^{-2}$, and $10^{-1}$, respectively, at $t=85$~years. All of these values are comparable to the result of \cite{2022Lykou}.

\begin{figure}
\includegraphics[width=1\columnwidth]{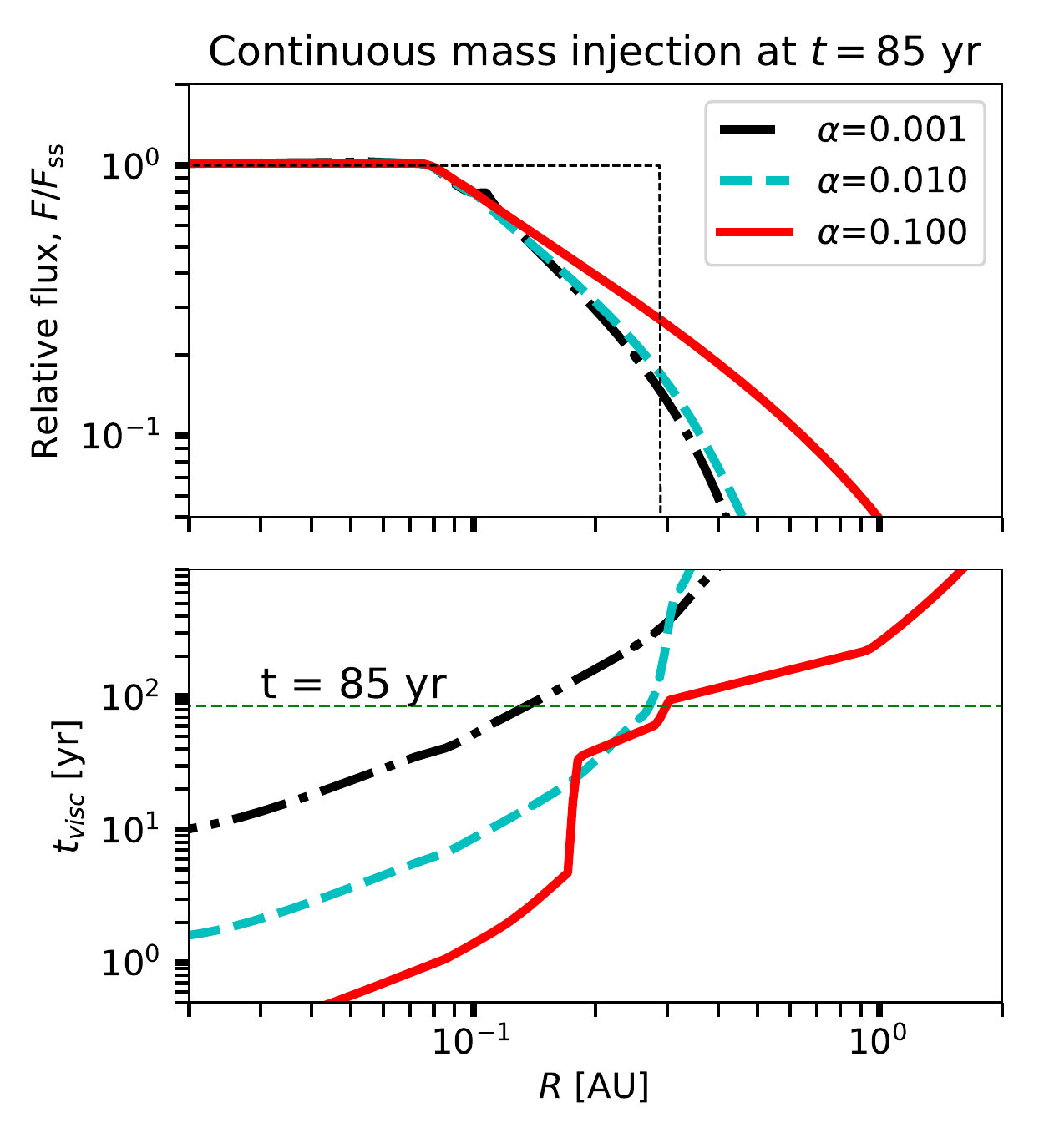}
\caption{Comparison of relative disc fluxes (eq. \ref{F_ss}) and the viscous time vs radius for the different values of $\alpha$ in the steady-state mass source model (\S \ref{sec:ad_hoc_steady}.) Note that for all $\alpha$ the size of the active disc is $0.2-0.3$ AU, close to the observed value.}
\label{fig:Fluxes_3alphas}
\end{figure}

The insensitivity of $R_{\rm ac}$ to the value of $\alpha$ stems from the fact that $h$ in eq. \ref{R_ac_planet0} actually anti-correlates with $\alpha$.  Fig. \ref{fig:Fluxes_3alphas} shows the relative radiative fluxes of the three  $\alpha$ models in the top panel, and the local disc viscous time in the bottom panel at $t=85$~years, the current age of FU Ori outburst. For all the values of $\alpha$ the viscous time at $R \ge 0.3$~AU is longer than 85 years. Therefore, the material outflowing to larger radii has not had time to reach beyond 0.3 AU. 



\subsection{Cumulative disc fluxes in L band}\label{sec:Radial_flux_distribution}

\begin{figure}
\includegraphics[width=1\columnwidth]{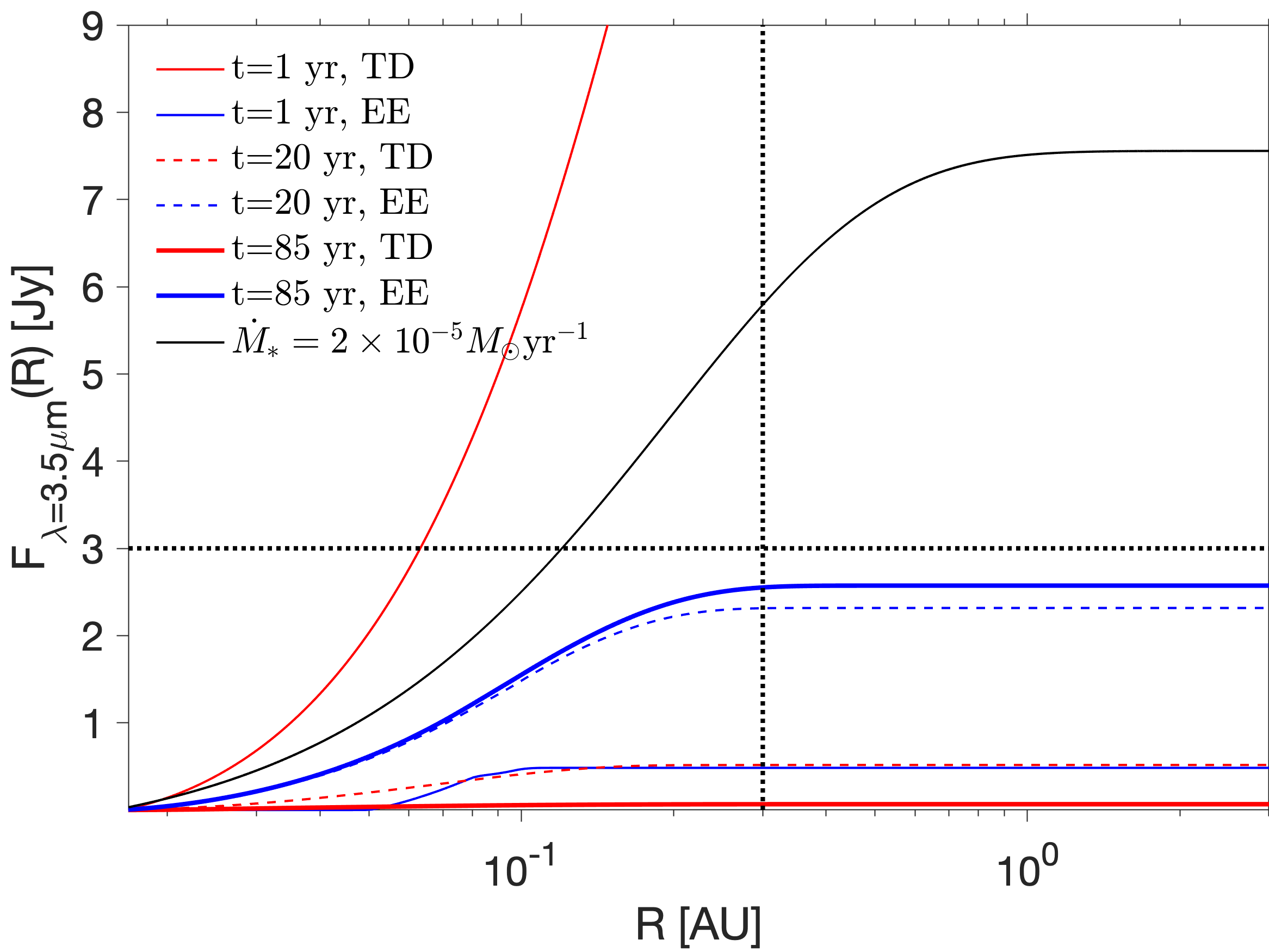}
\caption{Radial dependence of cumulative radiative flux in the models presented in \S\ref{sec:prompt_td} (red lines), \S\ref{sec:ad_hoc_steady} (blue lines), and in a steady state disc model with a constant stellar mass accretion rate $\dot{M}_{*} = 2\times 10^{-5}\msun$~year$^{-1}$ (black line). The vertical dotted line shows the active region size in FU Ori disc, $R=0.3$~AU. The horizontal dotted line shows the observed disc radiative fluxes at 3.5~$\mu m$ for FU Ori. \citep{2022Lykou}}
\label{fig:fluxes_lin}
\end{figure}

\cite{2022Lykou} marginally resolve FU Ori in the L band, concluding that the emission in FU Ori is neither a point source nor a region with $R$ much larger than 0.3 AU. They find that about $1/4$ of the flux in the L band comes from the passive disc at $R > 0.3$~AU. In this paper we do not model this reprocessing, and so we focus here just on the L band emission by the active disc. In Fig.~\ref{fig:fluxes_lin}, the observed radiative flux of actively accreting region in FU Ori disc at 3.5~$\mu m$ \citep{2022Lykou} is shown with the horizontal dotted line. For simplicity we assume the disc to emit as a local blackbody, then calculate the local emitted flux at L band, and compute the cumulative disc flux in this band at the observer as a function of the radius in the disc. In Fig.~\ref{fig:fluxes_lin}, cumulative disc fluxes are shown for the following models: instantaneous mass release (tidal disruption, TD; red lines, \S\ref{sec:prompt_td}), the steady state mass loss (steady state evaporation, EE; blue lines, see \S\ref{sec:ad_hoc_steady}) by the planet orbiting at $R=0.08$~AU; and a steady-state disc with mass accretion rate $\dot{M} = 2\times 10^{-5}\msun$~year$^{-1}$ (black line). The model fluxes are shown at different times.  The vertical dotted line shows the inferred active disc size in FU Ori \citep{2022Lykou}. The cumulative flux shown in the figure must saturate at the intersection of vertical and horizontal dotted lines for the model fluxes to match the observations.

The cumulative L band flux of the steady state model is higher than observed by a factor of 2.5. The radius of flux saturation in this model is  a factor of 3 larger than observed $R_{\rm ac}$. The model disc is too bright and too large\footnote{After this paper was accepted to publication, \cite{Bourdarot-23-FUOR} presented interferrometric observations of FU Ori in H and K NIR bands that also find an active disc radius of $\sim 0.3$~AU. While these results confirm \cite{2022Lykou} findings in the L band, interestingly the authors find that an MRI activation based scenario \citep[in the spirit of][]{ArmitageEtal01} fit the disc size quite well. This to some degree disagrees with our conclusions here. We intend to investigate this disagreement further in future work.}.

In the TD model, the flux spikes dramatically at early times when the cumulative flux is about an order of magnitude too high. During the next $\sim$30 years, the flux decreases rapidly. The flux saturation radius in the model at $t=30$~years is ~0.3--0.4~AU, close to the size of active disc size in FU Ori. However, the flux stays close to the observed values for only a short period of time and then continues to plummet, becoming by a factor of 50 too low (thick red line) at $t=85$~years. Thus, the model is able to reproduce the fluxes and active disc size for only a few years. It also contradicts the observed weak evolution of B and V band fluxes \citep[e.g.,][]{Clarke05-FUORs}. 

In the EE model, the radiative flux gradually increases after the planet starts losing mass. The growth slows down over time and reaches 2.6~Jy at $t=85$~years, being consistent with the observed value. Flux saturation is at $R\sim0.3$~AU. Unlike the TD model, disc flux in the EE model spends many decades close to the observed fluxes of FU Ori. 


\section{Why no TI in FU Ori?}\label{sec:no_TI}

The FU Ori lightcurve has been steadily declining over almost 100 years. While TI cannot provide a good match to the data, should we not expect TI instability to operate and make FU Ori vary strongly and non-monotonically over this long a time scale? The physics of TI applies to any accretion disc system exceeding the minimum accretion rate $\dot M_{\rm min}$  \citep{Bell94,Lasota01-Review} for instability. Importantly, $\dot M_{\rm min}$ is independent of the value of $\alpha$ because it is set by the ionisation conditions of Hydrogen \citep{Hameury-20-review}. 
\cite{Lasota08-S-curve} show that for a star of mass $m_* \msun$ with a Solar composition disc, TI operates inside a radius in the disc $R_{-1} =  R/(0.1$~AU) if $\dot M_{\rm feed}$ exceeds 
\begin{equation}
    \dot M_{\rm min} = 1.7\times 10^{-5} \frac{\msun}{\text{year}}\; R_{-1}^{2.58} m_*^{-0.85}\;.
    \label{dotM_min}
\end{equation}
Currently $\dot M$ in FU Ori is a few$\times 10^{-5}\msun$~year$^{-1}$ \citep{2022Lykou}, and historical data indicate it may have been a factor of $2-3$ higher in the past \citep[e.g.,][]{Clarke05-FUORs}. Based on eq. \ref{dotM_min}, we expect the inner $\sim 0.2$~AU of the disc in FU Ori to be unstable to TI.  $\dot M$  through the inner disc of FU Ori {\em should have} varied  by orders of magnitude during the past 85 years, but it has not.

A clue is provided by the experiments in \S \ref{sec:ad_hoc_steady} with an ad hoc steady-state mass source located at 0.08 AU. We found in \S \ref{sec:ad_hoc_steady} that, whatever the value of $\alpha$, the accretion rate onto the star is steady despite the planet injecting the mass into the disc at a rate $\sim 2\times 10^{-5}\msun$~year$^{-1}$. This $\dot M_{\rm feed}$ makes a conventional outside-in fed disc unstable. Evidently, feeding the disc from inside the unstable region may render it steady.

To explore this further, we run a series of numerical experiments set up similarly to those in \S \ref{sec:ad_hoc_steady} but for a range of source locations. In parallel to that, we employ the following prescription for the behaviour of $\alpha$ with the disc temperature:
\begin{equation}
    \ln\alpha = \ln\alpha_{\rm cold}\, + \,\frac{\ln\alpha_{\rm hot}-\ln\alpha_{\rm cold}}{1 + (T_{\rm cr}/T)^8}
    \label{alpha_vs_T}
\end{equation}
where $T_{\rm cr}=2.5\times 10^4$~K is a critical temperature. This anszat, with $\alpha_{\rm cold} = 0.01$ and $\alpha_{\rm hot} = 0.1$, was proposed by \cite{Hameury98-alpha} and was found to work relatively well for both observations \citep[e.g.,][]{Lasota01-Review} and first-principle simulations of TI in discs \citep{Hirose14-Convection,Hirose15}.


We begin with a disc in steady-state at all radii inside the computational domain, transporting the matter towards the star at a very low accretion rate, $\dot M = 10^{-9}\msun$~year$^{-1}$. A source of matter losing mass at rate $\dot M_{\rm p} = 2\times 10^{-5}\msun$~year$^{-1}$ is then placed into the disc at radius $a_{\rm s}$, where $a_{\rm s}$ is a parameter that ranges from $0.06$~AU to 1.2~AU for different runs.  As we aim to reproduce a quasi-steady accretion rate observed in FU Ori, we run the simulations long enough for the matter lost by the source to reach the star plus a few thousand years. This ensures that the time average accretion rate onto the star by the end of these experiments is close to $\dot M_{\rm feed}$.

\begin{figure}
\includegraphics[width=0.45\textwidth]{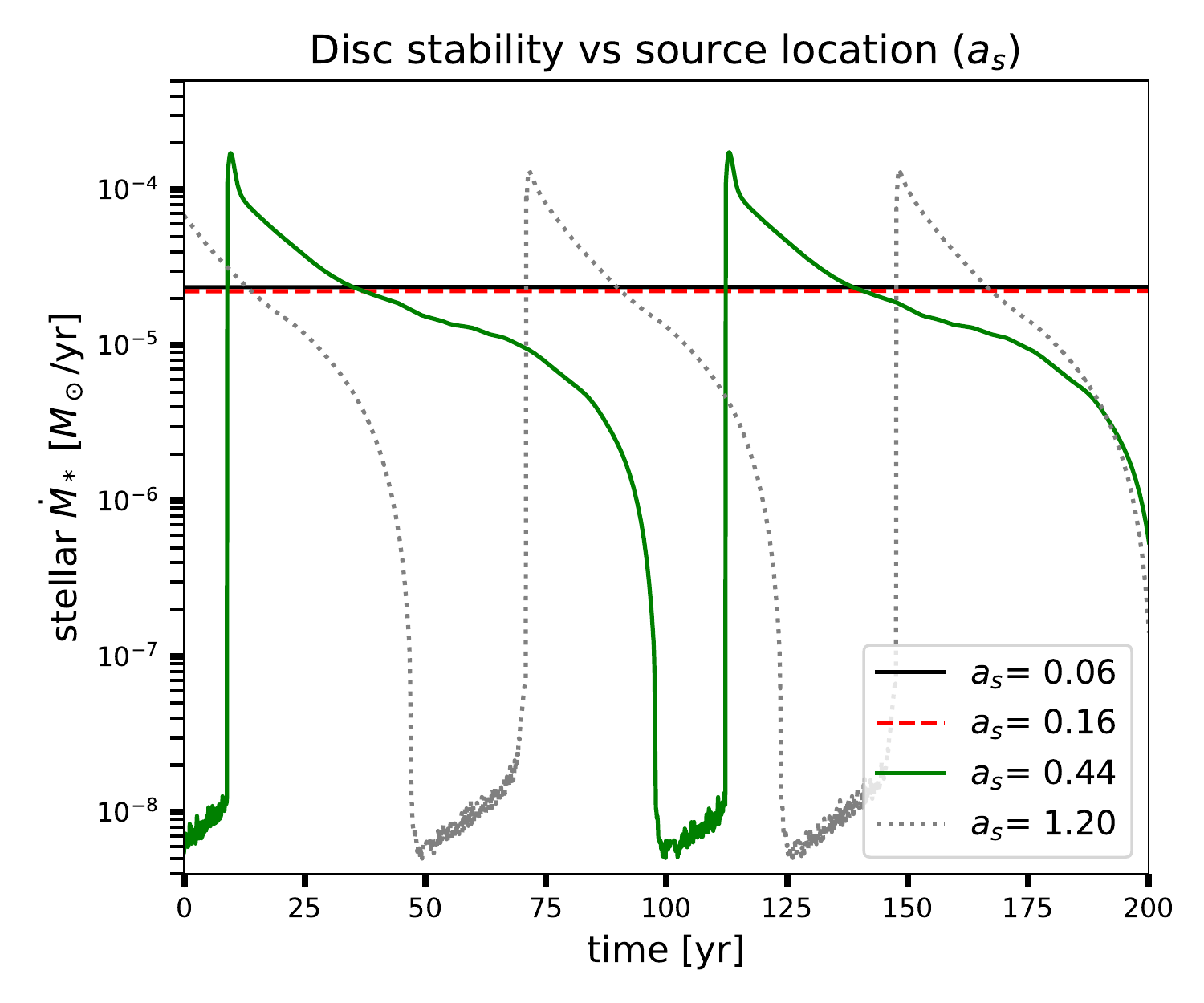}
\caption{Stellar accretion rate $\dot M_*$ versus time for experiments with an ad hoc steady-state source placed at different radial locations $a_{\rm s}$ as shown in the legend. Note that only when the source is located sufficiently close to the star $\dot M_*$ is steady.}
\label{fig:Mdot_vs_a_NO_TI}
\end{figure}

\begin{figure}
\includegraphics[width=0.45\textwidth]{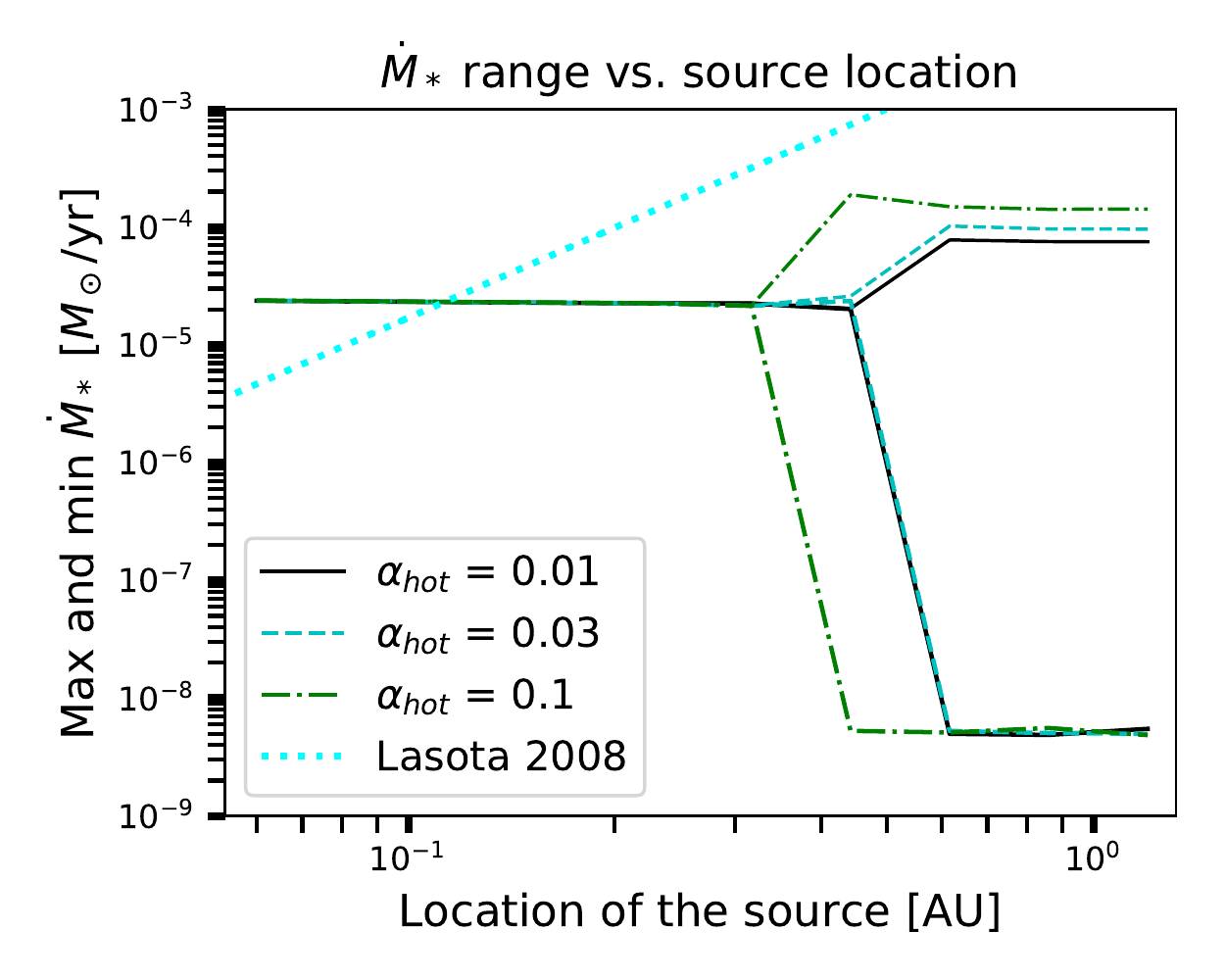}
\caption{The maximum and minimum accretion rate onto the star (cf. Fig. \ref{fig:Mdot_vs_a_NO_TI}) versus the source of matter location. To explain the long-term stability of FU Ori, the source must be located inside $\sim 0.3$~AU. See text in \S \ref{sec:no_TI} for detail. }
\label{fig:NO_TI}
\end{figure}

Fig. \ref{fig:Mdot_vs_a_NO_TI} shows stellar $\dot M$ versus time for such experiments, with the respective value of $a_{\rm s}$ shown in the legend. For $a_{\rm s} = 0.06$~AU and $0.16$~AU, accretion is steady, whereas for $a_{\rm s} = 0.44$~AU and $1.2$~AU TI is clearly present; we can rule such models out. To characterise the variability properties of such experiments versus $a_{\rm s}$, we measure the minimum and maximum accretion rates onto the star and plot these in Fig. \ref{fig:NO_TI}. To evaluate the dependence of the results on the $\alpha$ prescription, we also considered two other values of $\alpha_{\rm hot} = 0.01$ (same as $\alpha_{\rm cold}$), and $\alpha_{\rm hot} = 0.03$, and plot the results in Fig. \ref{fig:NO_TI}. The figure shows that disc accretion onto the star is in steady-state if the source of matter is located at $a_{\rm s} \lesssim 0.3$~AU, although there is a weak dependence on the $\alpha$ prescription. We also show in Fig. \ref{fig:NO_TI} the critical accretion rate $\dot M_{\rm min}$ given by eq. \ref{dotM_min}. If our disc behaved as the local disc S-curves predict \citep{Lasota08-S-curve} then we would expect $\dot M_*$ to be stable only for $a_{\rm s} \lesssim 0.12$~AU rather than $a_{\rm s} \sim 0.3$~AU. It is possible that our one vertical zone approximation is insufficiently accurate compared to the more detailed vertical balance models of \cite{Lasota08-S-curve}. Alternatively, the inner disc injection of mass and the non-local terms present in our modelling may be important. In any event, these experiments show that a quasi-steady source of matter placed inside the region $R \lesssim (0.1-0.3)$~AU from the star results in a steady $\dot M_*$. On the other hand, if the matter is introduced into the disc at larger distances, then we should expect FU Ori to have shown variability in $\dot M_*$ by many magnitudes since the beginning of the outburst, but
it has been remarkably steady over $\sim 85$ years \citep{Clarke05-FUORs,2022Lykou}. 

The absence of TI in FU Ori challenges the MRI activation scenario directly because the ``source" of the matter input into the active disc in this model is the inner edge of the dead zone, i.e., $\approx 2-3$~AU (\S \ref{sec:R_disc_MRI}). 2D simulations of the FU Ori hot disc by \cite{Zhu09-FUOR-2D-MRI-sims} confirm this conclusion. Such simulations are very expensive numerically when they cover more than one decade in radius, and so one often sets the inner region to exclude the TI-unstable zone. However, \cite{Zhu09-FUOR-2D-MRI-sims} found that when the inner $\sim 0.1$~AU of the disc is resolved in the simulations then TI does appear and leads to a significant and rapid variability in $\dot M$ on the star.

\section{Extreme evaporation of planets}\label{sec:EE}


\subsection{Preliminaries: thermal boil off}\label{sec:Parker}

\cite{OwenWu-16-boil-off}  studied thermally driven Parker winds from $\sim (1-10)$ Earth-mass planets exposed to irradiation from the central star. The two essential radial scales for the problem are the planet radius $R_{\rm p}$ and
the Bondi radius,
\begin{equation}
    R_{\rm B} = \frac{G M_{\rm p}}{2 c_s^2} \approx 21 R_{\rm J} \, m_3 T_4^{-1}\;,
\end{equation}
where $T_4 = T_{\rm h}/(10^4 $K) where $T_{\rm h}$ is the temperature of the hot thermal bath surrounding the planet, and $m_3 = M_{\rm p}/(3 \mj)$. As we will assume that the outer layers of the planet are warmed up to $T_{\rm h}$, $R_{\rm B}$ is where the planet's gas is no longer thermally bound to it and can escape it.


Simple estimates \citep[Fig. 1 in][]{Nayakshin10c}, semi-analytical models \citep[e.g.,the right panel in Fig. 4 of][]{HumphriesEtal19}, and 3D numerical simulations \citep[e.g.,][]{VB06,BoleyEtal10,FletcherEtal19} show that gas giants born at $\sim 50$~AU by gravitational instability may migrate into the inner AU in $\sim 10^4$ years. The radius of such planets can be as large as $R_{\rm p}\sim O(10) R_{\rm J}$; cf. \S \ref{sec:planet_origin} for detail.

Let us consider the case when $R_{\rm p}$ is smaller than the Hill radius $R_{\rm H}$ of the planet, $R_{\rm H} = a \left(\frac{M_{\rm p}}{3 M_*}\right)^{1/3} \approx 25 R_{\rm J} \; a_{-1} m_3^{1/3}$, where  $a_{-1} = a/(0.1$ AU), and we assumed $M_* = 0.6\msun$. \cite{OwenWu-16-boil-off} show that in the ``boil-off" regime $R_{\rm p} \ll R_{\rm B}$, the mass outflow rate is
\begin{equation}
    \dot M_{\rm p} = 4\pi R_{\rm p}^2 \rho_{\rm surf} v_{\rm out}\;,
    \label{Mdot_boild_off_1} \sim 6\times 10^{-8}\, \frac{\msun}{\hbox{yr}} \,m_3 T_4^{-1/2} \frac{v_{\rm out}}{c_{\rm h}} \kappa_{\rm R}^{-1} \,,
\end{equation}
where $\rho_{\rm surf}$ is the photospheric density, $\kappa_{\rm R}$ is Rosseland opacity, and $v_{\rm out}$ is strongly subsonic with respect to the sound speed  $c_{\rm h} = (k_b T_h/\mu)^{1/2} \approx 10\, T_4^{1/2}$~km/s. While very important for super-Earth planets, such mass loss rates are very small in terms of FU Ori phenomenon. 

\subsection{Optically thick energy limited outflow}\label{sec:ee_regime}

In isolation, photospheric temperatures of the youngest giant planets and brown dwarfs, $T_{\rm ph}$, is in the range $\sim (1-3)\times 10^3$~K \citep{Bodenheimer74,BodenheimerEtal80,BurrowsEtal97}. Consider the situation where the temperature of the external medium in which the planet is bathed suddenly increases by a large factor to $T_{\rm h} > 10^4$~K and the planet finds itself in the regime $R_{\rm B} < R_{\rm p} < R_{\rm H}$. As $ T_{\rm h} \gg T_{\rm ph}$, radiation diffuses  {\em into} the planet, heating its outer layers to $\sim T_{\rm h}$, making them unbound and launching an outflow.

We seek a steady-state solution in which the mechanical luminosity of the outflow is equal to the radiation energy flux entering the planet. As the flow is optically thick, radiation and gas are in thermodynamic equilibrium. In the outflow region, the gas temperature is a monotonically increasing function of radius $R$ counted in this section from the planet centre, and the radiation flux, $F_{\rm rad}$ is directed into the planet. 

Outflows from stars and planets are often supersonic at large distances. Here, however, the flow is unlikely to be supersonic. If it were supersonic at large distances, there would be a shock where the outflow ram pressure meets sufficient resistance from the surrounding disc. The shock would not be isothermal as the disc is optically thick. The gas would be heated to temperature $T \gg T_{\rm h}$. But then the radiation flux at that point would change sign and be positive. The outflow would be cut off from its ultimate energy source, the hot disc surrounding the planet, and would therefore shut down.

The steady-state energy equation for an evaporative flow is \citep[cf.][]{CowieMcKee-77}:
\begin{equation}
    {\vec \nabla} \cdot  \rho \vec{v} \left(\frac{5}{2} c_{\rm s}^2 + \frac{v^2}{2}\right) + {\vec \nabla} \cdot {\vec F_{\rm rad}} = 0
    \;,
    \label{Flow_master_eqn}
\end{equation}
where $c_{\rm s}$ and $v$ are the local sound and speed velocity of the gas, respectively. 

In spherical symmetry, 
\begin{equation}
    F_{\rm rad}(R) = - \frac{16}{3} \frac{\sigma_B T^3}{\kappa_{\rm R} \rho}\frac{d T}{dR}\;,
    \label{F_rad0}
\end{equation}
and equation \ref{Flow_master_eqn} can be integrated,
\begin{equation}
    \rho v \left(\frac{5}{2} c_{\rm s}^2 + \frac{v^2}{2}\right) = 
    \frac{16}{3} \frac{\sigma_B T^3}{\kappa_{\rm R} \rho}\frac{d T}{dR}\;.
    \label{Flow_master_eqn2}
\end{equation}
Here the constant of integration was set to zero due to boundary conditions at infinity: the radiation flux entering the planet from the disc (the right hand side of eq. \ref{Flow_master_eqn2}) is exactly equal to the outflow energy flux (the left hand side). Multiplying both sides of the equation by $4\pi R^2$ and identifying $\dot M_{\rm p} = 4\pi R^2 \rho v = $ const as an unknown eigenvalue of the problem, we have
\begin{equation}
    \dot M_{\rm p} \left(\frac{5}{2} c_{\rm s}^2 + \frac{v^2}{2}\right) = 
    \frac{64\pi}{3} \frac{\sigma_B R^2 T^3}{\kappa_{\rm R} \rho}\frac{d T}{dR}\;.
    \label{Flow_master_eqn3}
\end{equation}
We shall make a simplifying assumption that $v= c_{\rm s}$ everywhere in the flow. Since $R_{\rm p} > R_{\rm B}$ and the flow is thermally launched, it has enough thermal energy to escape everywhere, yet it is not likely to be accelerated much above $c_{\rm s}$ since it is thermally driven. The density $\rho$ is eliminated through $\rho = \dot M_{\rm p}/(4\pi R^2 c_{\rm s})$, and
\begin{equation}
    \dot M_{\rm p}^2 = 
    \frac{256\pi^2}{9} \frac{\sigma_B R^4 T^{5/2}}{\kappa_{\rm R}} \left(\frac{\mu}{k_b}\right)^{1/2} \frac{d T}{dR}\;.
    \label{Flow_master_eqn4}
\end{equation}
The opacity $\kappa_{\rm R}$ is a complicated function of density $\rho$ and temperature, and therefore eq. \ref{Flow_master_eqn4} cannot be solved analytically. 

\subsubsection{Order of magnitude estimate}\label{sec:mdot_energy_toy}

Let us first estimate $\dot M_{\rm p}$. The outflow is launched in the region close to the planet's surface, so by the order of magnitude $R \sim R_{\rm p}$, $dT/dR \sim (T_{\rm h}- T_{\rm ph})/R_{\rm p} \sim T_{\rm h}/R_{\rm p}$, and $T\sim T_{\rm h}$. With this, we obtain the ``characteristic" mass loss rate
\begin{equation}
        \dot M_{\rm char} = \frac{16\pi}{3} \left(\frac{\sigma_B}{\kappa_{\rm R}} \right)^{1/2} \left(\frac{\mu}{k_b}\right)^{1/4} R_{\rm p}^{3/2} T_{\rm h}^{7/4}\;.
     \label{Mdot_diff0}
\end{equation}
Numerically, with $\mu = 0.6 m_p$,
\begin{equation}
    \dot M_{\rm char} = 2.25\times 10^{-6}\; \frac{\msun}{\hbox{year}}\kappa_{\rm R}^{-1/2}  \left(\frac{R_{\rm p}}{10 R_J}\right)^{3/2} T_{\rm 4}^{7/4}\;.
\end{equation}
For $T_4 \sim$ a few, this is of the order of the accretion rate in the inner active disc of FU Ori. Note that Eq. \ref{Mdot_diff0} depends on opacity $\kappa_{\rm R}$, assumed constant throughout the flow in this estimate. Even in this case, to find $\dot M_{\rm p}$ at a given $T_{\rm h}$, one needs to solve eq. \ref{Mdot_diff0} for $\dot M_{\rm char}$ iteratively, by adjusting the estimate of $\rho = \dot M_{\rm char}/(4\pi R^2 c_{\rm s})$ and using this in opacity $\kappa_{\rm R}$. We find that in general $\kappa_{\rm R}\sim 10$ for the problem at hand.

\subsubsection{An approximate analytical solution}\label{sec:mdot_energy_solution}

In this section, we continue with our approximation that $\kappa_{\rm R} = $~const. This then allows us to integrate equation \ref{Flow_master_eqn4} in a closed form. The constant of integration follows from demanding $T(R_{\rm p}) = T_{\rm ph}$:
\begin{equation}
    \frac{512\pi^2}{3} \frac{\sigma_B}{\kappa_{\rm R}} \left(\frac{\mu}{k_b}\right)^{1/2}  (T^{7/2} -T_{\rm ph}^{7/2})= 
    \dot M_{\rm p}^2 \left( \frac{1}{R_{\rm p}^3} - \frac{1}{R^3}\right) 
    \label{Flow_solution1}
\end{equation}
The value of $\dot M_{\rm p}$ is now found by demanding $T(\infty) = T_{\rm h}$:
\begin{equation}
    \dot M_{\rm p} = 16\pi \sqrt{\frac{2}{3}} \left( \frac{\mu}{k_B}\right)^{1/4}\left(\frac{\sigma_B}{\kappa_{\rm R}} \right)^{1/2} 
     R_{\rm p}^{3/2} \left(T_{\rm h}^{7/4} - T_{\rm ph}^{7/4}\right)\;.
     \label{Mdot_analytic_1}
\end{equation}
If we neglect $T_{\rm ph}\ll T_{\rm h}$ in this equation then it is exactly $\sqrt{6} = 2.44$ times larger than $\dot M_{\rm char}$. Numerically,
\begin{equation}
     \dot M_{\rm p} = \dot M_{\rm ee} = 5.5 \times 10^{-6}\; \frac{\msun}{\hbox{year}}  \left(\frac{R_{\rm p}}{10 R_J}\right)^{3/2} \frac{T_{\rm 4}^{7/4}}{\kappa_{\rm R}^{1/2}}
     \left(1 - \frac{T_{\rm ph}^{7/4}}{T_{\rm h}^{7/4}}\right)\;.
     \label{Mdot_diff_ex2}
\end{equation}

\subsubsection{Radiation diffusion limiter on $\dot M_{\rm p}$}

Throughout the derivation in \S\S \ref{sec:mdot_energy_toy} \& \ref{sec:mdot_energy_solution}, we neglected the fact that radiation diffuses into the planet at a finite speed $c_{\rm diff}$. We can formally define this speed through $u_{\rm rad} c_{\rm diff} = |F_{\rm rad}|$, where $u_{\rm rad} = a_{\rm rad} T^4$ is the blackbody radiation energy density, obtaining
\begin{equation}
    c_{\rm diff} = c \;\frac{d\ln T^{4/3}}{d\tau}\approx \frac{c}{\tau_{\rm out}}\;.
\end{equation}
Here we used the fact that $T_{\rm h} \gg T_{\rm ph}$. We shall now recognise the fact that radiation is also advected outward with the flow at velocity $v_{\rm out}$. If $v_{\rm out} > c_{\rm diff}$, then external radiation will not reach the interior layers of the planet and the outflow will shut down. We must therefore require that $ c_{\rm diff}\gtrsim v_{\rm out}$, or equivalently, 
\begin{equation}
    \tau_{\rm out} = \frac{\kappa_{\rm R} \dot M_{\rm p}}{4\pi R_{\rm p} v_{\rm out}}\leq \frac{c}{v_{\rm out}}\;,
\end{equation}
which evidently results in an upper limit,
\begin{equation}
    \dot M_{\rm p} \leq \dot M_{\rm max} = \frac{4\pi c R_{\rm p}}{\kappa_{\rm R}}= 4.2\times 10^{-4} \; \frac{\msun}{\hbox{year}}\, \kappa_{\rm R}^{-1} \frac{R_{\rm p}}{10 R_{\rm J}}
    \;. \label{dotM_diffusion}
\end{equation}

More formally, and more accurately, a similar result is derived by including the radiation advection term into the left hand side of eq. \ref{Flow_master_eqn2}, which then becomes
\begin{equation}
       \dot M_{\rm p} \left(\frac{5}{2}c_{\rm s}^2 + \frac{v^2}{2} + \frac{E_{\rm rad}}{\rho}\right)  
\label{ee_rad}
\end{equation}
where the radiation energy density $E_{\rm rad} = a_r T^4$. When the outflow is radiation pressure dominated, the $E_{\rm rad}/\rho$ term in eq. \ref{ee_rad} is the dominant one. Upon neglecting the gas energy density terms one can derive the temperature structure of such a flow
\begin{equation}
    \ln T = \ln T_{\rm ph} + \frac{3 \kappa_{\rm R}}{16\pi c} \dot M_{\rm p} 
    \left( \frac{1}{R_{\rm p}} - \frac{1}{R}\right)
    \label{ee_rad_2}\;.
\end{equation}
Setting $T = T_{\rm h}$ at $R= \infty$, we have
\begin{equation}
    \dot M_{\rm diff} = \frac{16\pi c R_{\rm p}}{3\kappa_{\rm R}} \ln \left( \frac{T_{\rm h}}{T_{\rm ph}}\right)= \frac{5.6\times 10^{-4} }{\kappa_{\rm R}} \frac{\msun}{\hbox{year}} \frac{R_{\rm p}}{10 R_{\rm J}} \ln \left( \frac{T_{\rm h}}{T_{\rm ph}}\right)
    \label{ee_rad_3}
\end{equation}
As $\ln (T_{\rm h}/T_{\rm ph}) \sim 2-4$, this results in pre-factor in eq. \ref{ee_rad_3} of $\sim (1\div 2) \times 10^{-3} \msun$~year$^{-1}$. We see that the radiation diffusion limit to $\dot M_{\rm p}$ is usually significantly larger than eq. \ref{Mdot_diff_ex2}, unless both $T_4 \gg 1$ and $\kappa_{\rm R} \gg 1$.

\subsubsection{Numerical solution}\label{sec:numerical_mdot}

Lastly, we perform direct numerical integration of eq. \ref{Flow_master_eqn4}, not assuming now that $\kappa_{\rm R} =$~const, for a given set of problem parameters, that is, $M_{\rm p}$, $R_{\rm p}$, and $T_{\rm h}$. In doing so we first guess a value for $\dot M_{\rm p}$. We then eq. \ref{Flow_master_eqn4} outward in small increments of $R$ to ensure small changes in $T$ in every step, starting at $R=R_{\rm p}$, and setting $T(R_{\rm p}) = 4\times 10^3$~K. This temperature is still much lower than $T_{\rm h}$ and is large enough so that H$_2$ molecules would be largely dissociated. For a more accurate solution, one needs to include H$_2$ dissociation and H ionisation, but we expect only a minor correction since the gain in the specific internal energy of gas as it warms up to $T_{\rm h} \sim $ a few $\times 10^4$~K far outweighs these additional energy sinks. Note that radiation flux drops as $1/R^2$ at large radii from the planet since $4\pi R^2 F(R)=$ const for a steady state solution, implying that $dT/dR \rightarrow 0$, and hence $T(R)$ tends to a constant ($T(\infty)$) at $R\rightarrow \infty$. In general $T(\infty) \ne T_{\rm h}$; we then iterated on the value of $\dot M_{\rm p}$ until we matched the outer boundary condition, $T(\infty) = T_{\rm h}$ at $R \gg R_{\rm H}$. 

\begin{figure}
\includegraphics[width=1\columnwidth]{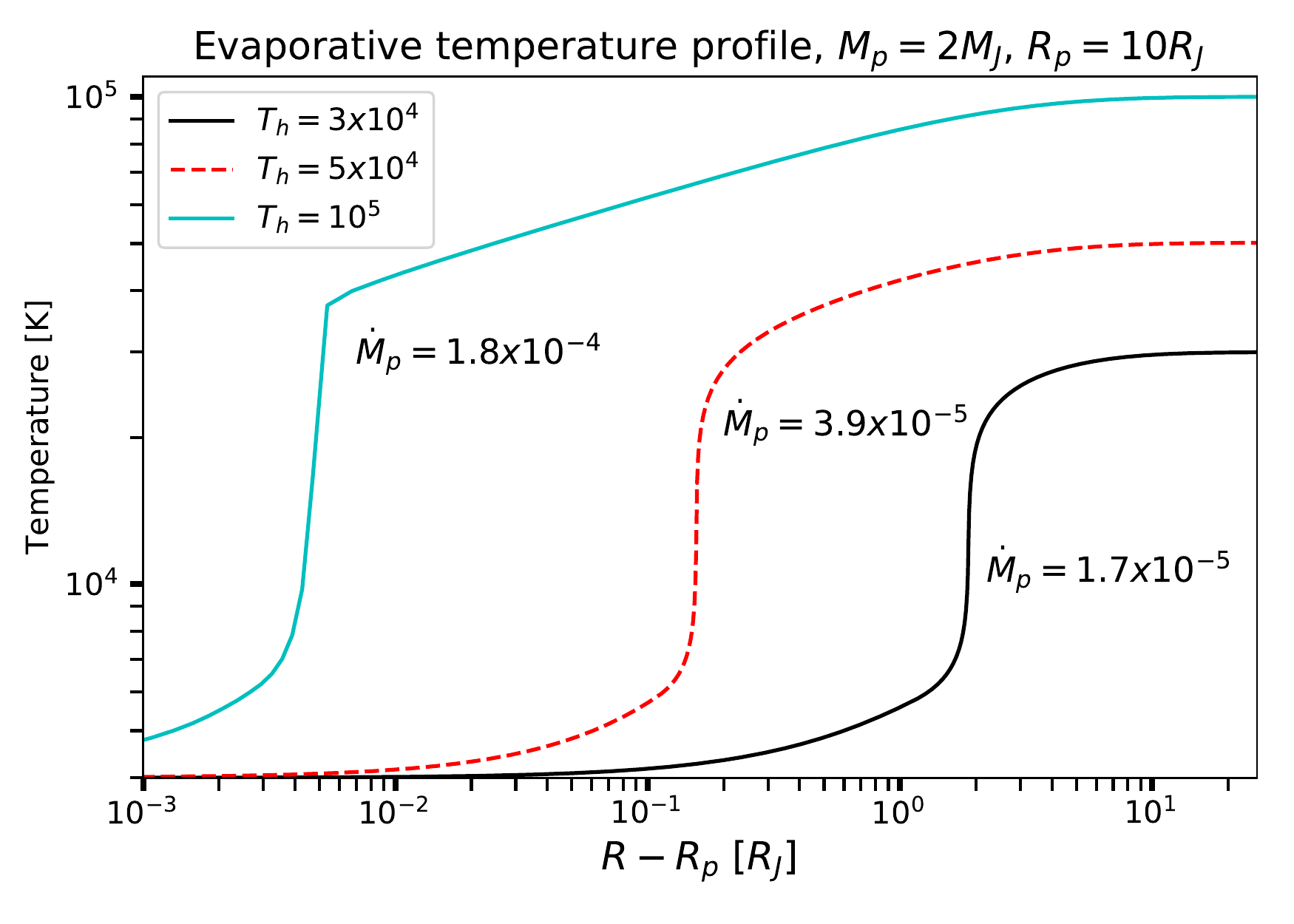}
\caption{Radial temperature profiles for the Extreme Evaporation of the planet computed as described in \S \ref{sec:numerical_mdot}. Three values of the hot bath temperature $T_{\rm h}$ are considered.}
\label{fig:T_vs_R_profiles}
\end{figure}

Fig. \ref{fig:T_vs_R_profiles} shows $T(R)$ profiles for $R_{\rm p}= 10 R_J$, $M_{\rm p} = 2\mj$ and three different values of $T_{\rm h}$. We can see that the larger $T_{\rm h}$, the steeper the temperature rise from the planet's surface. The width of the radial zone where the temperature suddenly increases to above $10^4$ K is always much smaller than a fraction of $R_J$, which is very small compared to the planet's radius. Fig. \ref{fig:T_vs_R_profiles} also  shows the planet mass loss rates derived through this numerical procedure. Encouragingly, these values are within $\sim 20$\% of the approximate $\dot M_{\rm p}$ given by eq. \ref{Mdot_diff_ex2} (we remind the reader that one still needs to iterate on $\kappa_{\rm R}$ in that equation to find an accurate expression for the mass loss rate, although $\kappa_{\rm R}\approx 10$ usually gives a result accurate to within a factor of 2).

\begin{figure*}
\includegraphics[width=1\columnwidth]{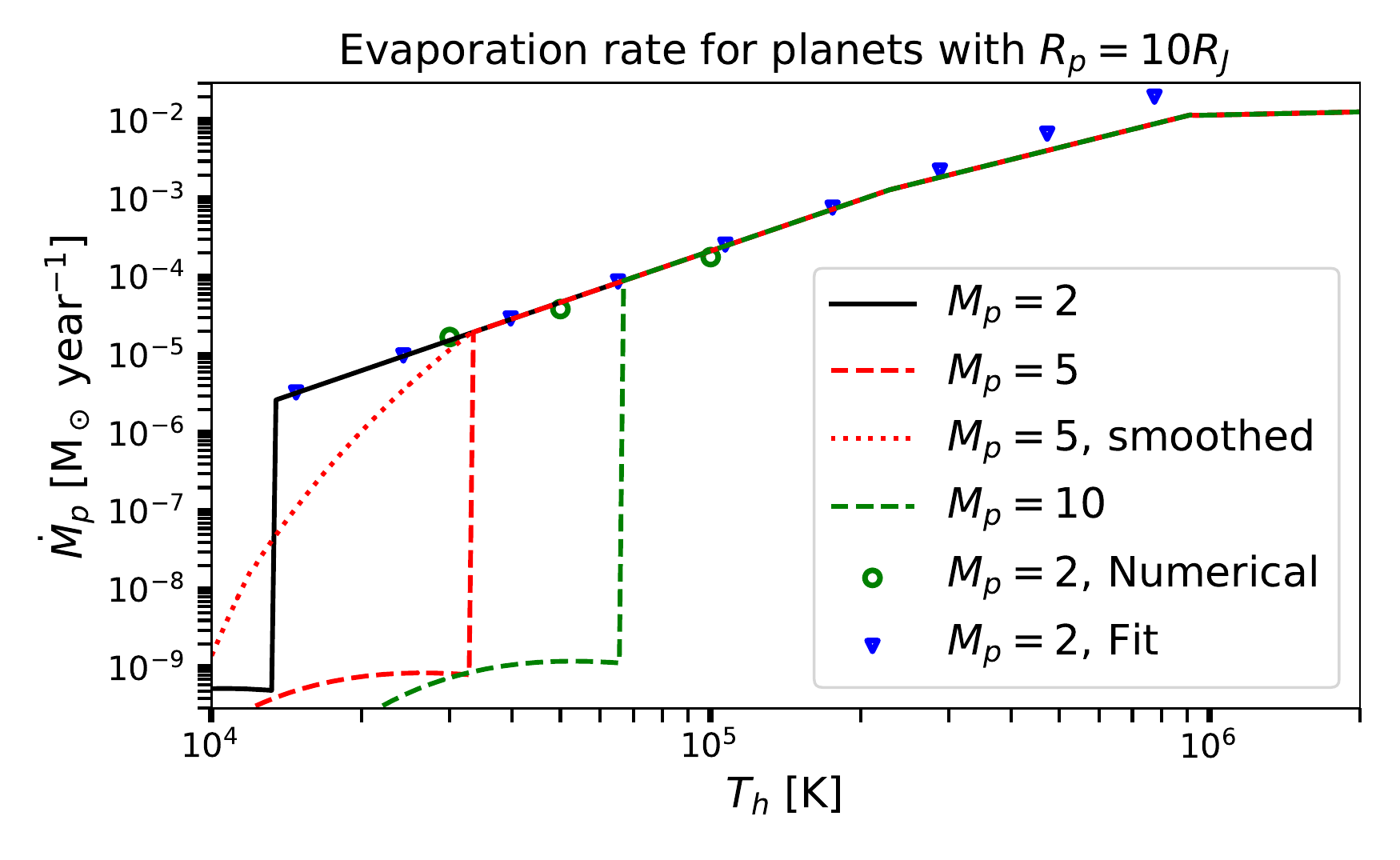}
\includegraphics[width=1\columnwidth]{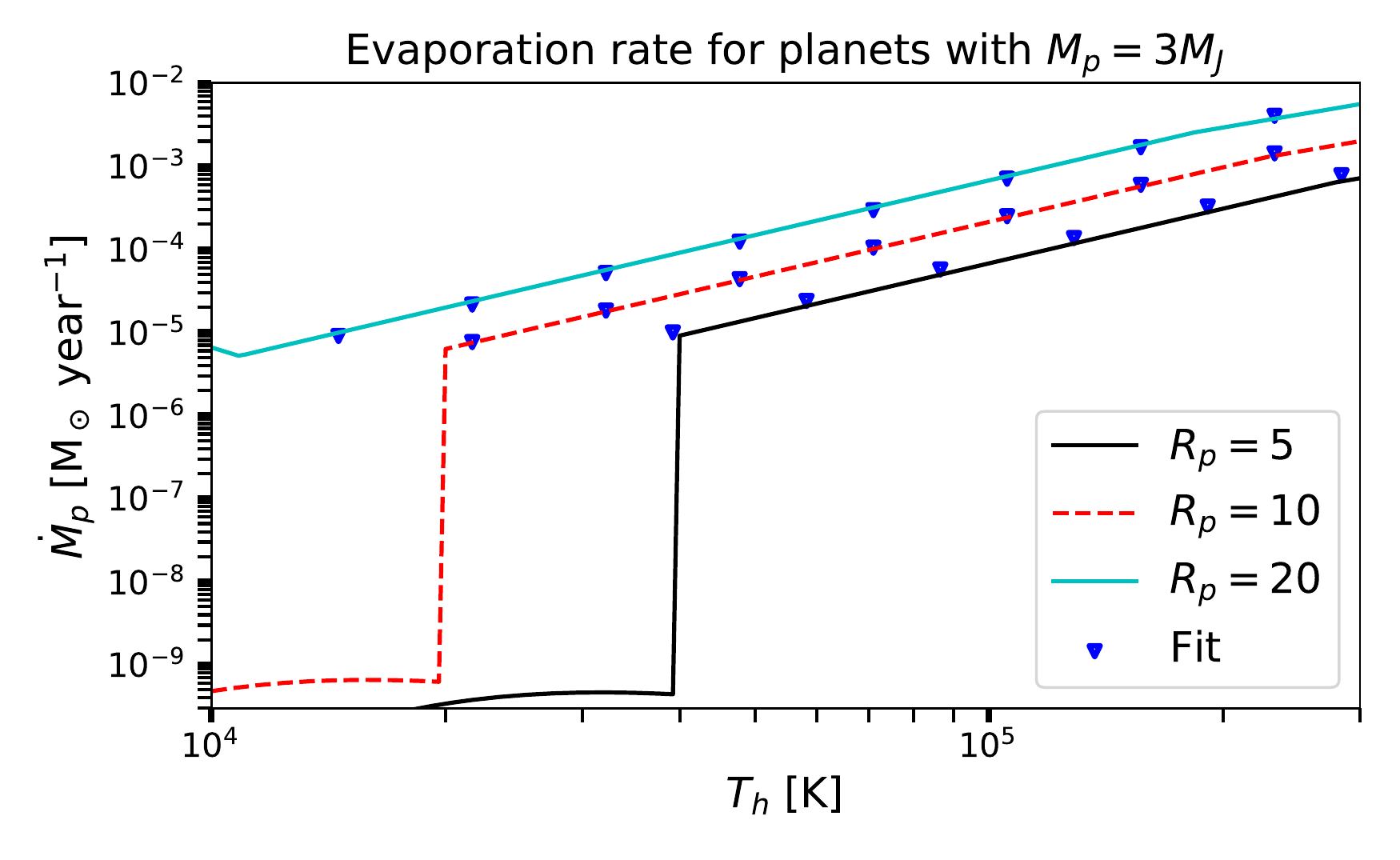}
\caption{Dependence of planet Extreme Evaporation rates computed via eqs. \ref{Mdot_boild_off_1} \& \ref{Mdot_diff_ex2} on the surrounding gas temperature for a selection of planet parameters. Left: Fixed planet radius, $R_{\rm p} = 10 R_J$, and varying $M_{\rm p}$. The smoothed evaporation model is discussed in \S \ref{sec:Ignition_smoothed}. The green circles and blue triangles show the numerical solution (\S \ref{sec:numerical_mdot}) and the approximation via eq. \ref{Mdot_fit}, respectively. Right: same but for a fixed $M_{\rm p}$ and varying $R_{\rm p}$.}
\label{fig:Mdot_vs_T}
\end{figure*}

Fig. \ref{fig:Mdot_vs_T} shows $\dot M_{\rm p}$ obtained from eq. \ref{Mdot_diff_ex2} versus $T_{\rm h}$ for a selection of planet parameters. In the left panel of the figure, we show planets with $R_{\rm p} = 10 R_J$ but with three different values of $M_{\rm p}$. Note that since we neglected planet gravity (recall that $R > R_{\rm p} > R_{\rm B}$ so the gas is not bound to the planet), there is no dependence of $\dot M_{\rm p}$ on $M_{\rm p}$, except at relatively low $T_{\rm h}$. In particular, if $T_{\rm h}$ is low, $R_{\rm B} > R_{\rm p}$, and so extreme evaporation does not take place. In that case, we use the Parker wind solution which yields much lower values of $\dot M_{\rm p} \sim 10^{-9} \msun$~year$^{-1}$. The three green circles in the left panel of Fig. \ref{fig:Mdot_vs_T} correspond to the numerical solutions obtained for $M_{\rm p} =2 \mj$ and plotted in Fig. \ref{fig:T_vs_R_profiles}. Therefore, in the rest of the paper, we shall use eq. \ref{Mdot_diff_ex2} for a numerically quick evaluation of $\dot M_{\rm p}$. For analytical purposes we also note the following numerical fit that works quite well:
\begin{equation}
    \dot M_{\rm p} = 1.4\times 10^{-6} 
    \frac{\msun}{\hbox{year}} \left(\frac{T_{\rm h}}{10^4 \hbox{K}} \right)^{\xi_p} \left(\frac{R_{\rm p}}{10 R_{\rm J}}\right)^{3/2}\;,
    \label{Mdot_fit}
\end{equation}
(blue triangles in Fig. \ref{fig:Mdot_vs_T}) where $\xi_p = 2.2$, until the radiation diffusion limit (eq. \ref{dotM_diffusion}) is reached.



\section{Idealised disc-planet co-evolution}\label{sec:disc-planet}


In this section we start with idealised experiments that inject the planet into the inner disc and neglects its radial migration. In this section the planet ``feels" the surrounding disc's heat but not its tidal torque; the disc also feels no torques from the planet but receives the mass lost by the planet (in terms of eq. \ref{dSigma_dt}, the second term on the right is set to zero but the last term is on). Unrealistically we also keep $M_{\rm p}$=const in this section; this is relaxed in \S \ref{sec:complete}.

\subsection{FUOR ignition, and two system modes}\label{sec:ignition_standard}

When do planet-disc systems produce FUOR-like behaviour? To investigate this we start with a fixed $\alpha = 5\times 10^{-3}$ planet-free disc 
fed at $\dot M_{\rm feed} =  10^{-7} \msun$~year$^{-1}$ at infinity. At this $\dot M_{\rm feed}$ the disc switches between quiescent periods with $\dot M_{\rm q}\sim 10^{-8} \msun$~year$^{-1}$ lasting $\sim 70$~years, and outbursts with $\dot M = \dot M_{\rm b} \sim 10^{-6} \msun$~year$^{-1}$ lasting $\sim 10$~years. We then inject a planet with mass $M_{\rm p} = 3\mj$ into the disc at $R= 0.08$  after 300 years of its planet-free evolution.

 
In quiescence, $T \lesssim 2\times 10^3$ K, and planet evaporation is negligible. However, at the peak of TI outbursts the EE rate may be comparable to or even exceed $\dot M_{\rm b}$. We shall now see that if $\dot M_{\rm p} < \dot M_{\rm b}$ then the planet is in the ``disc-controlled" mode, in which the planet simply follows the local disc conditions as it goes through the TI cycle. In the opposite case, if during a TI outburst $\dot M_{\rm p} > \dot M_{\rm b}$, then the planet-disc system becomes highly non-linear. The planet becomes the main source of matter in the inner disc, exceeding the mass supply from the outside planetary orbit. We call this system mode ``planet-sourced". As mass is released by the planet, the disc heats up, so $\dot M_{\rm p}$ increases further; an FUOR-like outburst with accretion rate $\approx \dot M_{\rm p} \gg \dot M_{\rm b}$ is ignited. 

\begin{figure*}
\includegraphics[width=1\columnwidth]{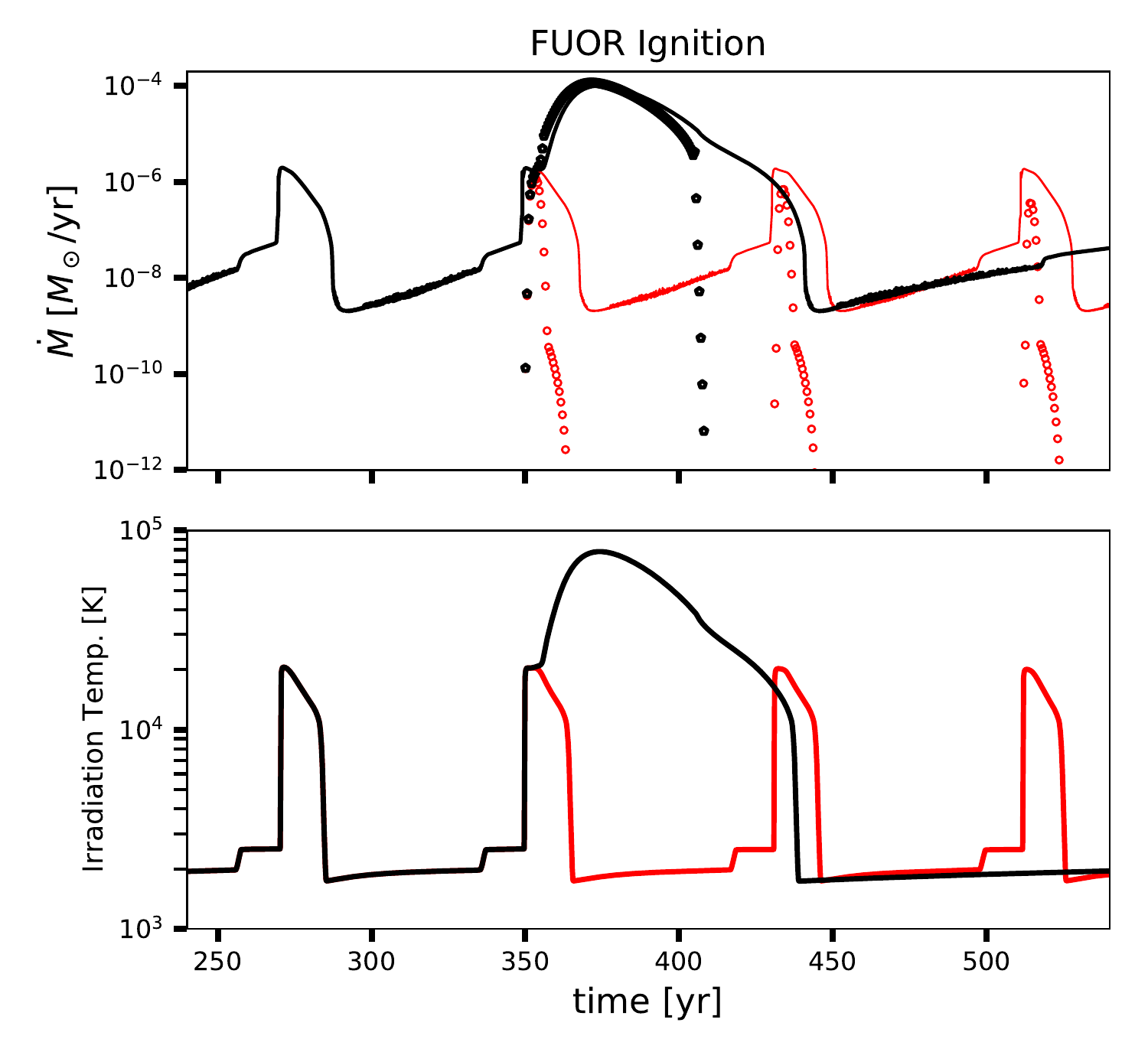}
\includegraphics[width=1\columnwidth]{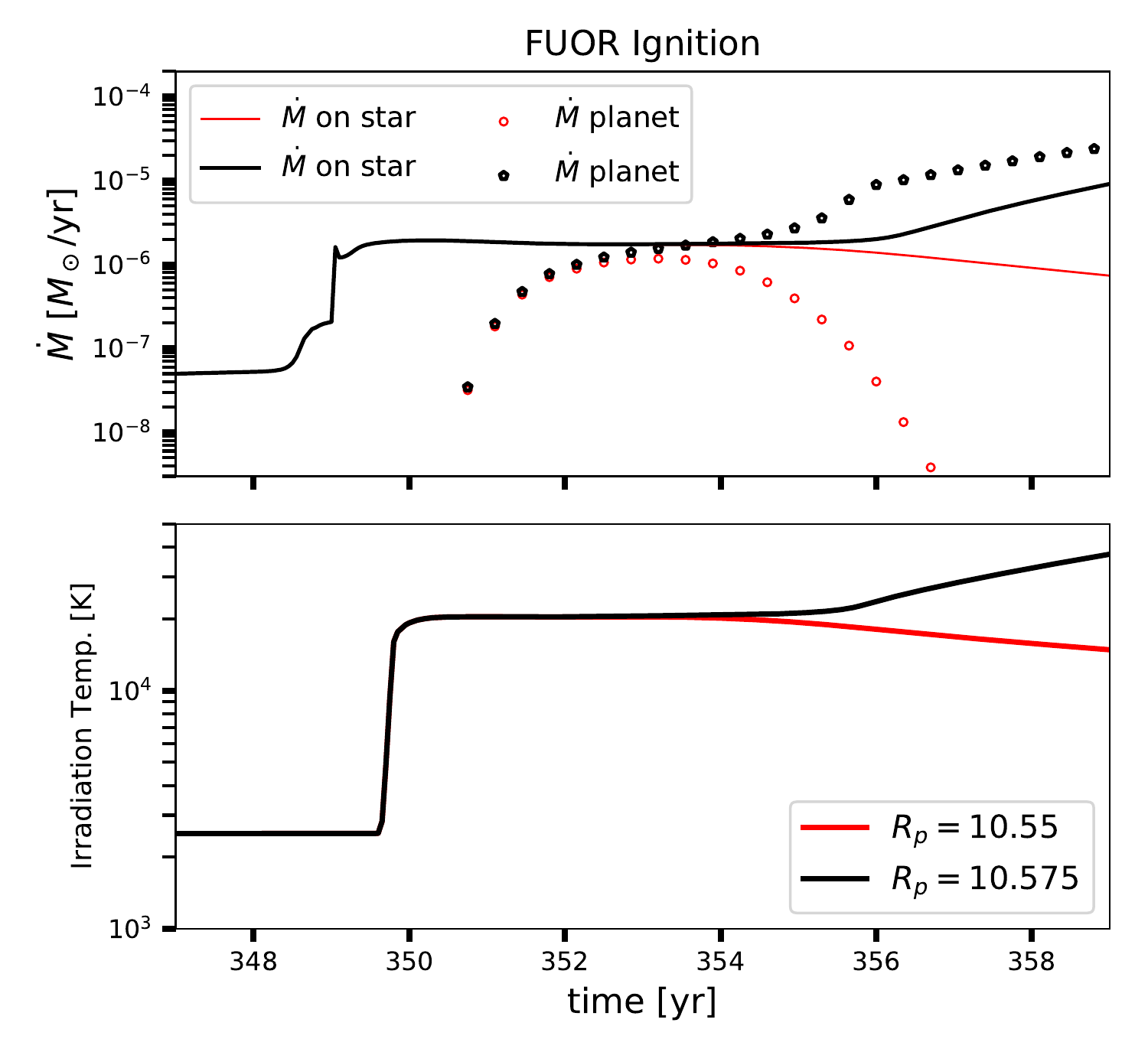}
\caption{Idealised experiments with a planet orbiting the star at a fixed radius of 0.08 AU (\S \ref{sec:ignition_standard}) Note that the FUOR outburst ignites (black) when $\dot M_{\rm p} \gtrsim \dot M_*$ for a period of time, and fails to ignite (red curves) otherwise. }
\label{fig:Ignition}
\end{figure*}

To delineate the two modes of the system we perform experiments for a range of $R_{\rm p}$, finding that there exists a critical value, $R_{\rm cr}$. For the parameters chosen, $R_{\rm cr} \approx 10.57 R_J$. Fig. \ref{fig:Ignition} shows the accretion rate onto the star (solid curves) and the planet mass loss rates, $\dot M_{\rm p}$ (circles) in the top panels, and the disc temperature (called planet irradiation temperature in the panels) at the location of the planet in the bottom panels. The left and right panels show the same curves, but the right panels zoom in on the time when the red and black curves diverge. For the former curves, $R_{\rm p}$ is just smaller $R_{\rm cr}$, and for the latter $R_{\rm p} > R_{\rm cr}$. Comparing the top and the bottom panels in Fig. \ref{fig:Ignition} we observe a positive feedback loop in the system in the case $R_{\rm p} > R_{\rm cr}$. When $\dot M_{\rm p}$ exceeds $\dot M$, the disc heats up, so the planet is irradiated with a higher temperature. This leads to a higher $\dot M_{\rm p}$, and then both increase with time above the values they would have in the disc without the planet.

Note that $R_{\rm p}$ is just a parameter controlling $\dot M_{\rm p}$ and taking the system across the FUOR ignition condition $\dot M_{\rm p} \gtrsim \dot M_{\rm b}$ in this section for the disc model chosen here. In a more general setting, $\dot M_{\rm b}$ depends on system parameters, e.g., $\alpha$, $M_*$, and the location of the planet. There is then no fixed critical radius $R_{\rm p}$ but a range for it. A robust condition for the master mode/FUOR ignition is thus not on the planet radius but rather on the planet mass loss rate, $\dot M_{\rm p} \gtrsim \dot M_{\rm b}$, whatever the latter may be.



\subsection{FUOR ignition: does a smoother transition in $\dot M_{\rm p}$ matter?}\label{sec:Ignition_smoothed}




In our  derivation of $\dot M_{\rm p}$ we broke the problem into two regimes, Parker wind ($R_{\rm B} \gg R_{\rm p}$, \S \ref{sec:Parker}) and extreme evaporation ($R_{\rm B} < R_{\rm p}$, \S \ref{sec:ee_regime}).  This results in a discontinuous jump in $\dot M_{\rm p}$ when going from one regime to the other. A more accurate solution is likely to result in  a smoother transition from Parker wind to EE. In \S \ref{sec:ignition_standard} we found that FUOR ignition is a surprisingly sharp condition in terms of $R_{\rm p}$; it was sufficient for $R_{\rm p}$ to increase by a fraction of 1\% above a critical value for the system to go into the self-sustained FUOR mode. One may worry that this behaviour is only due to the artificially sharp transition between the two planet mass loss regimes.

To gain an insight into how the main conclusions of our paper may depend on the idealised discontinuous transition in $\dot M_{\rm p}$ from the Parker wind to the EE model we introduce a  toy ``smoothed" extreme evaporation scenario in which we use EE $\dot M_{\rm ee}$ for $R_{\rm B} < R_{\rm p}$, but modify $\dot M_{\rm p}$ in the Parker regime:
\begin{equation}
    \dot M_{\rm p} = \dot M_{\rm se} = \begin{cases}
            \dot M_{\rm ee} \exp\left(-\frac{\Delta R}{\Delta R_{\rm tr}}\right)\; \text{ if } R_{\rm p} < R_{\rm B} \;,\\
            \\
        \dot M_{\rm ee}\; \text{ if } R_{\rm B} < R_{\rm p} 
    \end{cases}
    \label{Mdot_smoothed}
\end{equation}
Here $\Delta R = R_{\rm B} - R_{\rm p}$, and $\Delta R_{\rm tr} = \zeta R_{\rm p}$, a transition width parameter, with $\zeta = 0.25$ explored below. The respective planet mass loss rate is shown with the red dotted curve in the left panel of Fig. \ref{fig:Mdot_vs_T} for $M_{\rm p}=5\mj$ case. Comparing the red dotted and red dashed curves we observe a significantly smoother increase in $\dot M_{\rm p}$ with increasing $T_{\rm h}$ in the smoothed model. 



\begin{figure*}
\includegraphics[width=1\columnwidth]{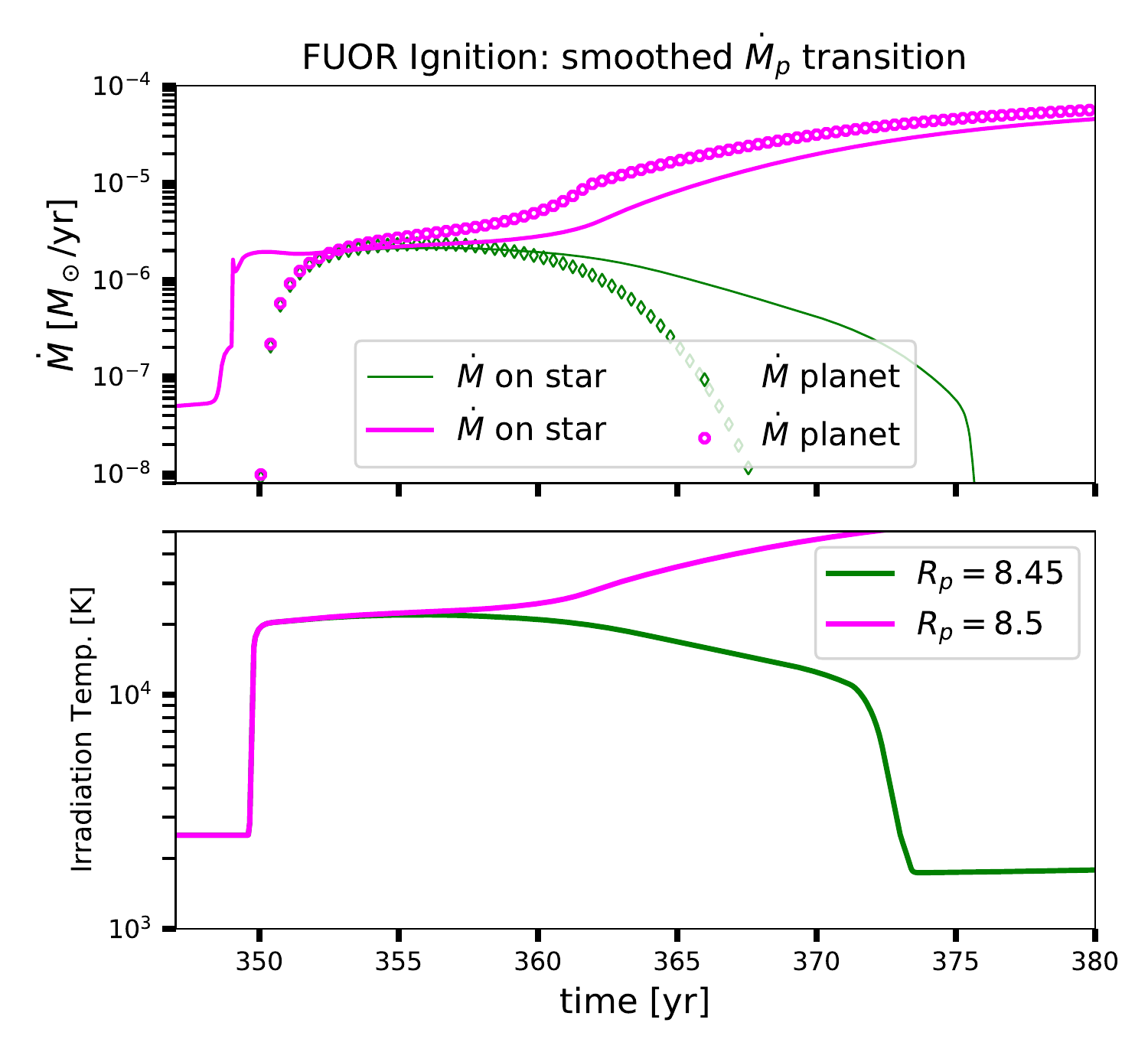}
\includegraphics[width=1\columnwidth]{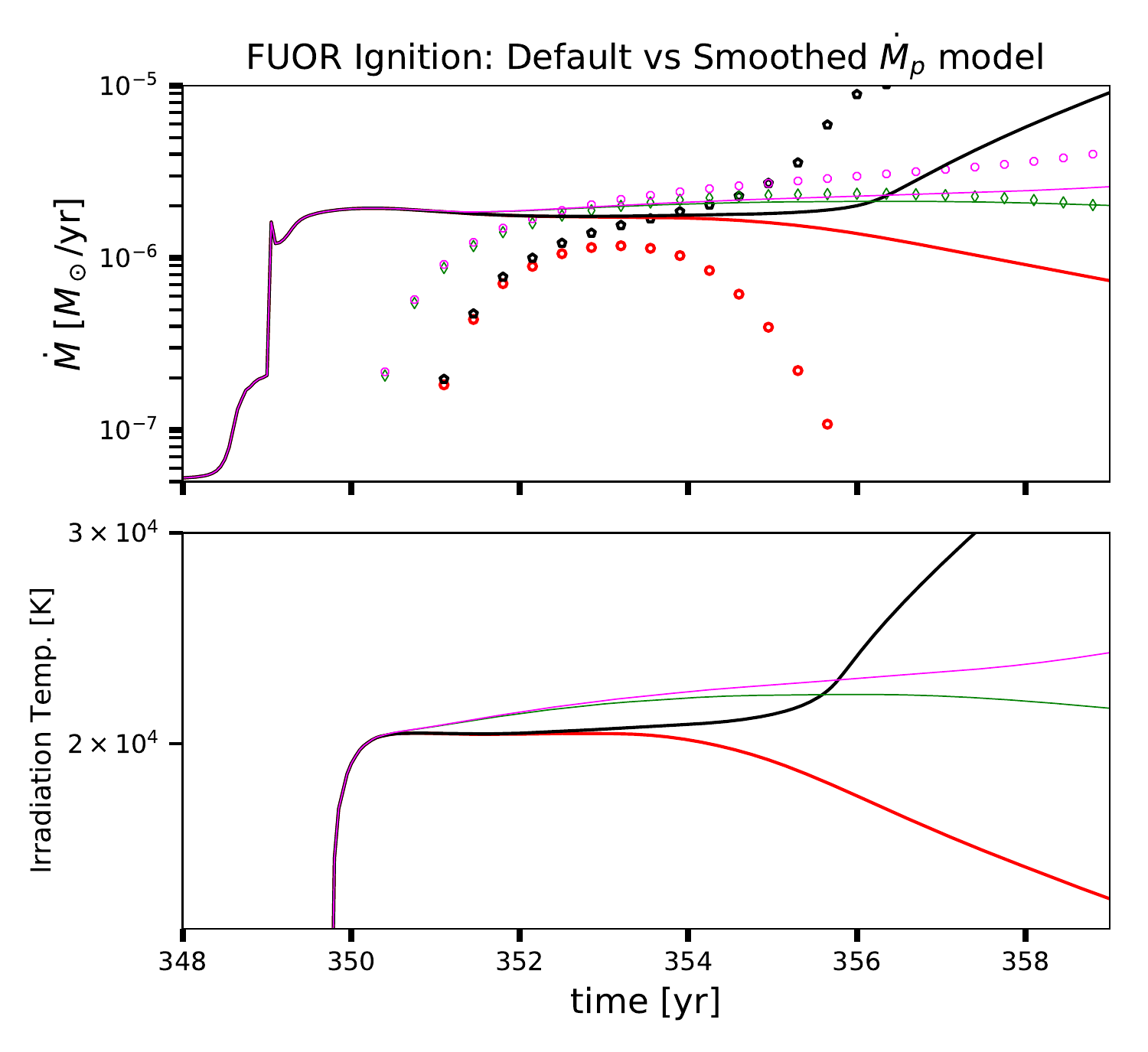}
\caption{Left: similar to Fig. \ref{fig:Ignition}, but for a smoothed transition into the extreme evaporation regime. Note that as in Fig. \ref{fig:Ignition}, FUOR ignition takes place above a critical $R_{\rm p}$ but not below it, albeit at smaller $R_{\rm p}$. Right: comparison of the standard (Fig. \ref{fig:Ignition}) scenario for $\dot M_{\rm p}$ with the smoothed evaporation case shown in the left panel. Note that smoothed model results in a smoother transition to FUOR ignition. See \S \ref{sec:Ignition_smoothed} for detail.} 
\label{fig:Ignition_smoothed}
\end{figure*}

Encouragingly for our model of FU Ori, we find that despite a much more gradual increase in $\dot M_{\rm p}$ with $T_{\rm h}$ in the smoothed model (eq. \ref{Mdot_smoothed}), the distinction between and a sharp transition between the self-sustained FUOR-like planet master and the planet slave modes remains. The left panel of Fig. \ref{fig:Ignition_smoothed} shows two experiments with $R_{\rm p} = 8.45 R_J$ and $R_{\rm p} = 8.5 R_J$. As in Fig. \ref{fig:Ignition}, there is a rather different behaviour of the system for the smaller and larger planet radii, although the critical planet radius is now smaller. This could be expected. For $\Delta R_{\rm tr} = 0.25 R_{\rm p}$ and the critical planet radius of $R_{\rm cr} \sim 10.5 R_J$ found in \S \ref{sec:ignition_standard}, $\Delta R_{\rm tr} \sim 2.5 R_J$. We can thus expect that the FUOR ignition condition in the smoothed $\dot M$ model sets in at smaller radii, at $R_{\rm p} \sim R_{\rm cr} - \Delta R_{\rm tr} \sim 8 \mj$. We also see that FUOR ignition condition is the one on the $\dot M_{\rm p}$: while the critical radius changed, the $\dot M_{\rm p}$ at which FUOR ignition occurs did not. The right panel of Fig. \ref{fig:Ignition_smoothed} compares the experiments with the standard and the smoothed $\dot M_{\rm p}$. The smoothed $\dot M_{\rm se}$ model results in a more gradual approach to the FUOR ignition or turn-off than the standard $\dot M_{\rm p}$ scenario. For example, $\dot M_{\rm p}$  approaches $\dot M_*$ earlier in the smoothed evaporation model by about a year. As a result, the disc temperature in the planet's vicinity increases earlier in the $\dot M_{\rm se}$ model. For the same reason, the sub-critical and critical models (green and magenta colours) diverge from each other slower than the red and the black ones do. 

These experiments show that (1) the existence of the self-sustained FUOR-like regime in the disc-planet system is independent of how rapidly the planet approaches EE regime; (2) Planets with smaller $R_{\rm p}$ are able to power FUOR-like disc activity in the smooth EE scenario; (3) The detail of FUOR light curves produced, however, depend on the sharpness of the transition between the Parker wind and the EE regime. 

\subsection{A steady state or runaway planet evaporation?}\label{sec:runaway}


Consider how the system evolves after the criterion $\dot M_{\rm p} \gtrsim \dot M$ is satisfied. For simplicity, we continue to set $R_{\rm p}$= const. Let the disc temperature be $T_{\rm ig}$ at the point of EE ignition. As the local disc surface density rises, so does $T_{\rm h}$, and thus $\dot M_{\rm p}$. Due to this growth in local $\Sigma$ and $T_{\rm h}$, the disc accretion rate $\dot M$ is increasing too. If $\dot M$ grows with $T_{\rm h}$ faster than does $\dot M_{\rm p}$, then a quasi-steady state equilibrium exists at some large $T_{\rm h} > T_{\rm ig}$ when $\dot M_{\rm p} \approx \dot M$. In that equilibrium, the disc is able to transfer the mass lost by the planet into the star. In the opposite case, if $\dot M$ grows with $T_{\rm h}$ slower than does $\dot M_{\rm p}$, then after the ignition $\dot M_{\rm p} > \dot M$ always, no matter how high $T_{\rm h}$ becomes. This will result in a runaway in $\dot M_{\rm p}$ and $\dot M$.


Eq. \ref{Mdot_fit} shows that $\dot M_{\rm p}\propto T_{\rm h}^{\xi_p}$ with $\xi_{\rm p} = 2.2$. For $T > T_{\rm ig}$, 
\begin{equation}
    \dot M_{\rm p} = \dot M_{\rm p}(T_{\rm ig}) \left(\frac{T}{T_{\rm ig}}\right)^{\xi_{\rm p}}\;.
\end{equation}
At the same time, steady-state one vertical zone disc equations can be solved for Kraemers' opacity $\kappa_{\rm R} = k_0 \rho T^{-7/2}$ to show that the disc midplane temperature $T \propto \dot M^{3/10}$, so
\begin{equation}
    \dot M= \dot M(T_{\rm ig}) \left(\frac{T}{T_{\rm ig}}\right)^{10/3} \;.
    \label{Mdot_ss73_vs_T}
\end{equation}
This shows that an equilibrium between $\dot M_{\rm p}$ and $\dot M$ should be possible at a temperature high enough since $\xi_{\rm p} < 10/3$. Conversely, if $\xi_{\rm p} > 10/3$, no equilibrium should be possible.

\begin{figure}
\includegraphics[width=1\columnwidth]{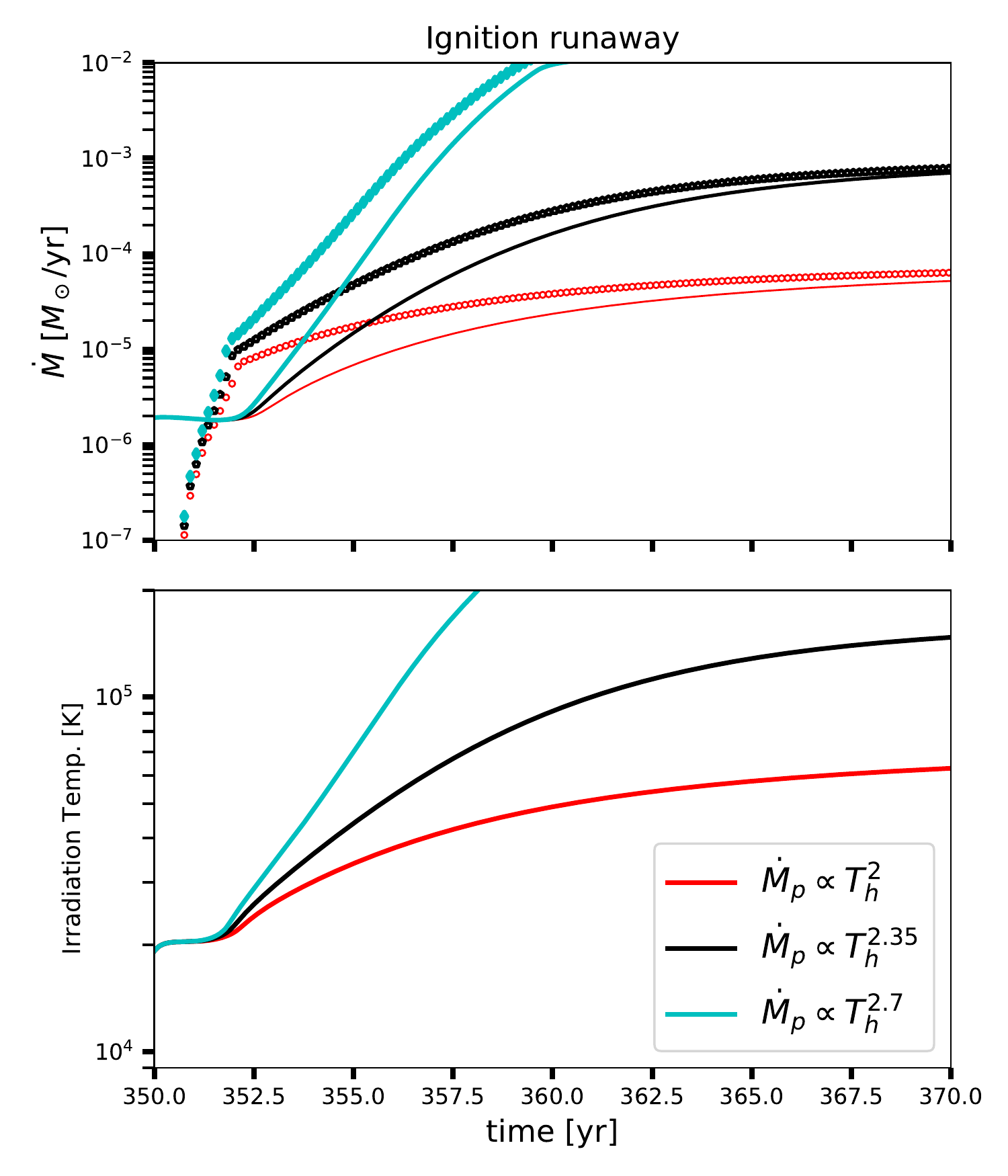}
\caption{Same experiments as shown in Fig. \ref{fig:Ignition}, but for a fixed $R_{\rm p} = 10.6 R_J$ and varying power-law indexes $\xi_p$ in the $\dot M_{\rm p}$ vs $T_{\rm h}$ dependence (eq. \ref{Mdot_fit}). Note that above $\xi_p \sim 2.5$, there is no quasi-steady state in the system; the disc keeps heating up too rapidly as $\dot M_{\rm p}$ increases, driving a further increase in $\dot M_{\rm p}$, leading to a runaway.} 
\label{fig:Ignition_runaway}
\end{figure}

\begin{figure}
\includegraphics[width=1\columnwidth]{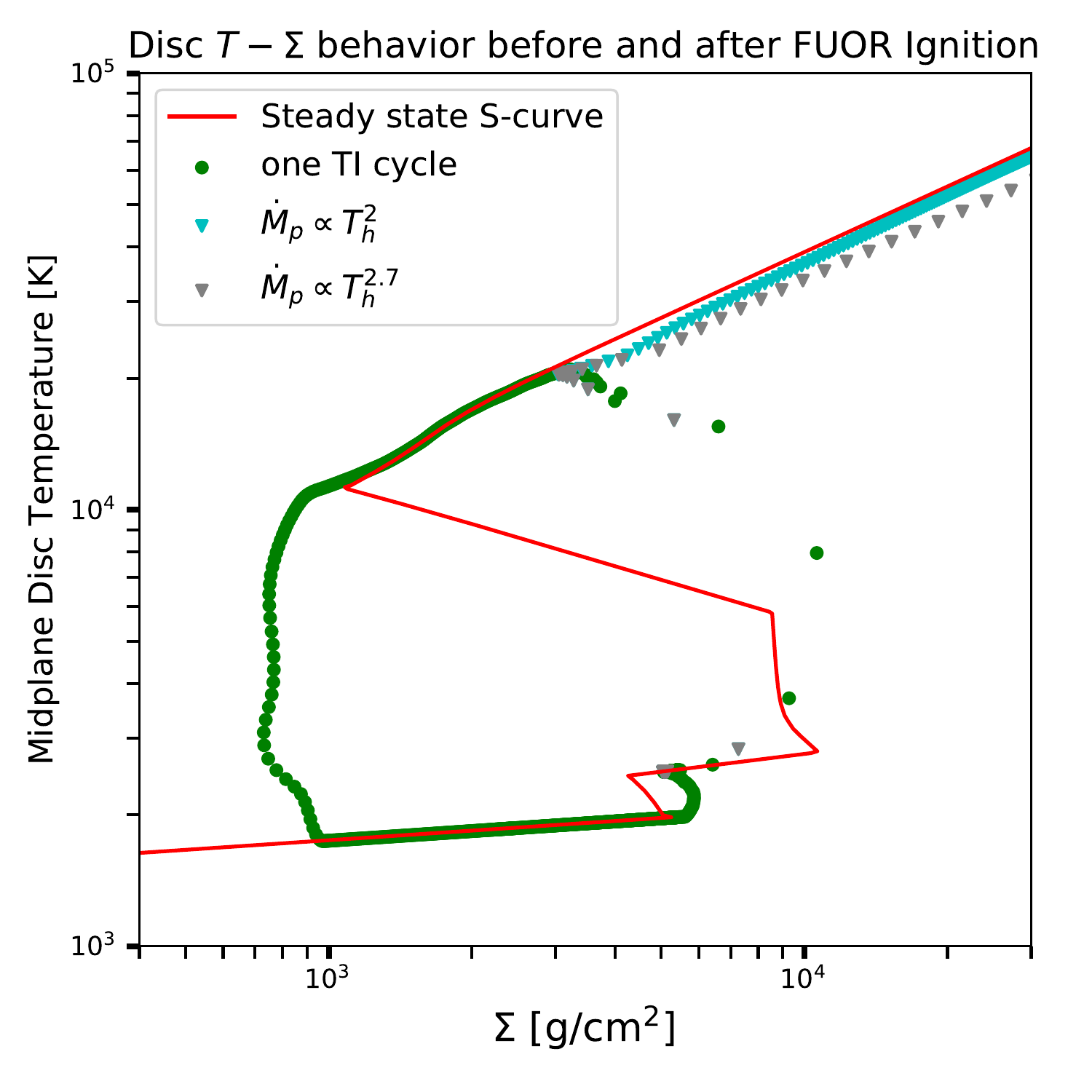}
\caption{The track of the disc midplane $T$ vs $\Sigma$ at the location of the planet for two experiments shown in Fig. \ref{fig:Ignition} is compared to the steady-state solution (the S-curve). The green circles show the disc evolution through one TI cycle before $\dot M_{\rm p}$ is turned on. The triangles show system evolution after EE ignition, when the ``unexpected" new mass supply increases local $\Sigma$ and pushes the disc up the S-curve. Note that the stronger $\dot M_{\rm p}$ grows with disc temperature, the stronger it deviates from the S-curve.} 
\label{fig:Ignition_S-curve}
\end{figure}

In Fig. \ref{fig:Ignition_runaway} we show numerical experiments set up in the same way as in \S \ref{sec:ignition_standard} (Fig. \ref{fig:Ignition}) except that we use a fixed $R_{\rm p }= 10.6 R_J$ and $\dot M_{\rm p}$ as per eq. \ref{Mdot_fit}  albeit with $\xi_{\rm p}$ different from 2.2 computed in \S \ref{sec:EE}. We observe that a quasi-steady state is reached for $\xi_{\rm p} = 2$ and $\xi_{\rm p} = 2.35$. For $\xi_{\rm p} = 2.7$ the system may be evolving towards a quasi-steady state but at such a high $\dot M_{\rm p}$ that many model assumptions start to break down (e.g., the disc becomes super Eddington and radiation pressure dominated). Fig. \ref{fig:Ignition_runaway} thus shows that in practice, for an equilibrium $\dot M_*$ in the observed FUOR range, $\xi_{\rm p}$ should be significantly lower than $10/3$.  Fig. \ref{fig:Ignition_S-curve} shows the evolution of the disc at the location of the planet for the experiments from Fig. \ref{fig:Ignition}. This phase portrait of the disc shows yet again that the planet takes the disc above the values of $T_{\rm d}$ and $\Sigma$ normally reachable in the TI cycles.

There are also constraints on the disc viscosity. The disc's viscous time at the location of the planet should be shorter than $t_{\rm p} = M_{\rm p}/\dot M_{\rm p}$, the mass loss time scale for the planet, or else the planet will lose all of its mass before the disc is able to transfer this mass onto the star. As $h = H/R \approx 0.2$ at $R=0.08$~AU,  $\alpha_{\rm h} \gtrsim 10^{-3} (100 \hbox{ yrs}/t_{\rm p})$.
FU Ori brightened on the time scale of $\sim 1$ year. Therefore, the disc's viscous time at the location of the planet should be no longer than a few years, that is, $\alpha_{\rm h} \gtrsim $~few$\times 10^{-2}$.

\subsection{A toy migrating planet example}\label{sec:results_example}


So far in this section, a planet was held on a fixed orbit in the inner disc. Real planets, of course,  must migrate into that region. From this section onward, we initiate planets far from the TI-unstable region and let them migrate radially due to tidal torques. The planet mass is evolved according to $dM_{\rm p}/dt = - \dot M_{\rm p}$. From this section the mass loss rate $\dot M_{\rm p}$ also includes the Roche Lobe overflow \citep[for details see \S 2.2 in][]{NayakshinLodato12}. The latter is small as long as $R_{\rm p}$ is smaller than the Hill radius, $R_{\rm H}$, but increases very rapidly when $R_{\rm p}$ approaches $R_{\rm H}$. What happens after FUOR ignition depends on the mass-radius relation for the planet. As it loses mass, it may expand or contract, and that may lead to significant changes in $\dot M_{\rm p}$. Here we assume a power-law form for the mass-radius relation,
\begin{equation}
    R_{\rm p} = R_{\rm p0} \left(\frac{M_{\rm p}}{M_{\rm p0}}\right)^{\xi_{\rm p}}\;,
    \label{M-R-relation}
\end{equation}
where $R_{\rm p0}$ and $M_{\rm p0}$ are the initial planet radius and mass before the mass loss sets in, and $\xi_{\rm p}$ is a dimensionless constant. For a polytropic sphere made of ionised Hydrogen with constant entropy $\xi_{\rm p} = -1/3$, but we explore a range of $\xi_{\rm p}$ below. 

\begin{figure*}
\includegraphics[width=0.99\textwidth]{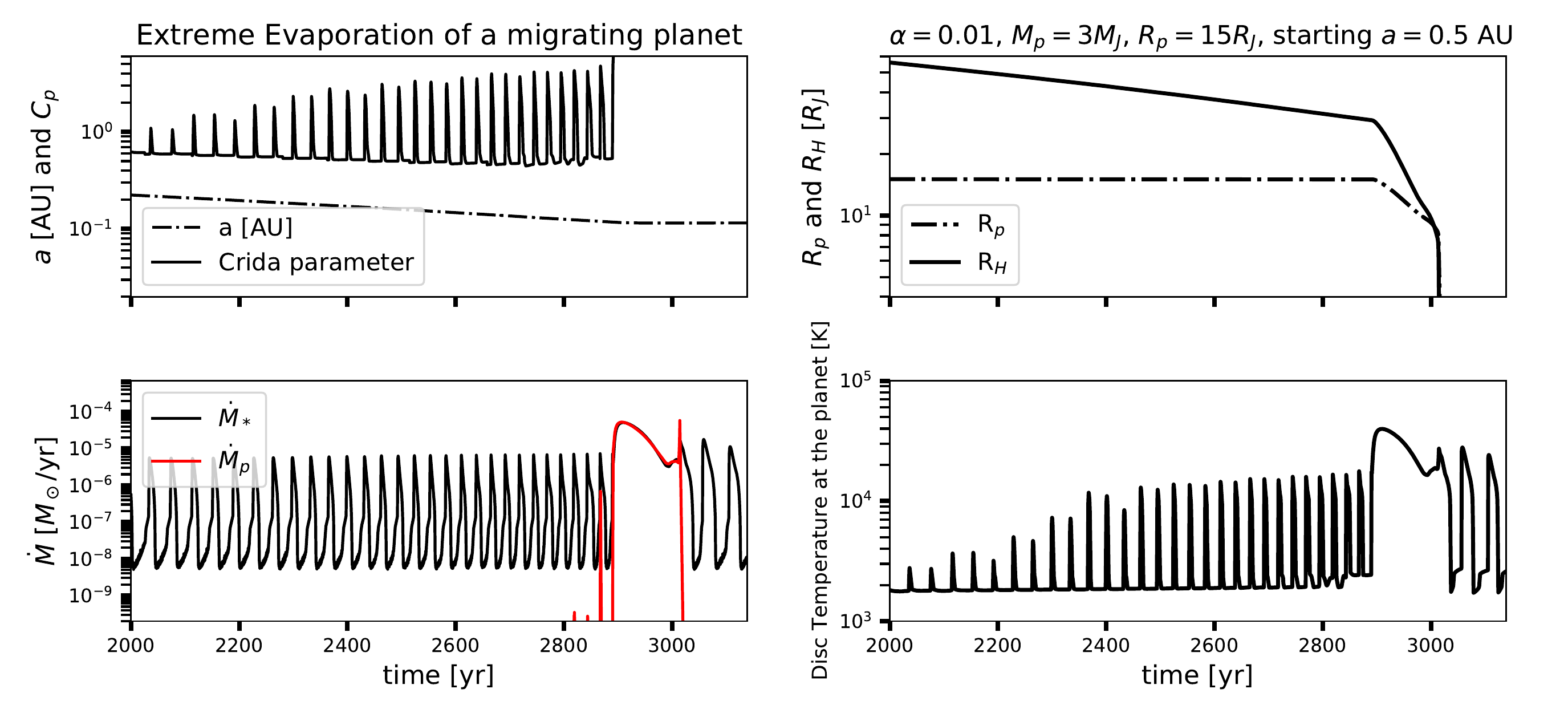}
\caption{Evolution of the disc-planet system. separation $a$, planet outflow rate, accretion rate onto the star, and the hot bath (irradiation) temperature for a planet starting in a disc at $a=0.3$~AU. }
\label{fig:a0.3_long}
\end{figure*}

Fig. \ref{fig:a0.3_long} shows evolution of the disc-planet system with a $M_{\rm p} = 3\mj$, $R_{\rm p0} = 15 R_{\rm J}$ planet, with $\xi_{\rm p} = 0.15$, inserted into the disc at $a=0.5$~AU. The planet's starting position is far enough to let the disc-planet system adjust into a self-consistent albeit non-linear interaction pattern by the time extreme evaporation occurs (or not). The disc viscosity parameter is fixed at $\alpha = 10^{-2}$. The disc is fed at a steady rate $\dot M_{\rm feed} =  10^{-6}\msun$~year$^{-1}$ at large distances.
For figure clarity, we do not show the first 2000 years of system evolution during which the planet migrated to $a\approx 0.2$~AU. 

The top left panel of Fig. \ref{fig:a0.3_long} shows the planet separation, $a$, and Crida parameter $C_{\rm p}$ \citep{CridaEtal06}. The planet initially migrates in the type II regime ($C_{\rm p}<1$, a deep gap in the disc is opened). This is the regime explored by \cite{LodatoClarke04}. However, their planets are more massive than ours, $M_{\rm p} \sim (10-15) \mj$ vs $M_{\rm p} = 3\mj$ in Fig. \ref{fig:a0.3_long}, and their viscosity is lower, $\alpha = 10^{-4}-10^{-3}$. Their planets were, therefore, always in the type II migration regime. In contrast, as our planet nears the inner TI-unstable disc,  it's $C_{\rm p}$ experiences excursions into the $C_{\rm p} > 1$ territory during TI flares. These excursions are due to the disc temperature (the bottom right panel of Fig. \ref{fig:a0.3_long}) increasing from $\lesssim 2\times 10^3$~K in quiescence to $\gtrsim 10^4$~K in TI outburst. $R_{\rm H}$ is shown in the top right panel of Fig. \ref{fig:a0.3_long}. The planet is  safe from tidal disruption, $R_{\rm p} \ll R_{\rm H}$, but it suffers EE at $a\approx 0.11$~AU. The bottom left panel shows $\dot M_{\rm p}$ with the red line. While the planet is far from the TI-unstable region, $T_{\rm h} \approx 2000$~K, $\dot M_{\rm p}$ is negligibly small. At $t = 2900$~years, the ignition condition is satisfied, and the system enters the planet-sourced regime.


\begin{figure}
\includegraphics[width=0.49\textwidth]{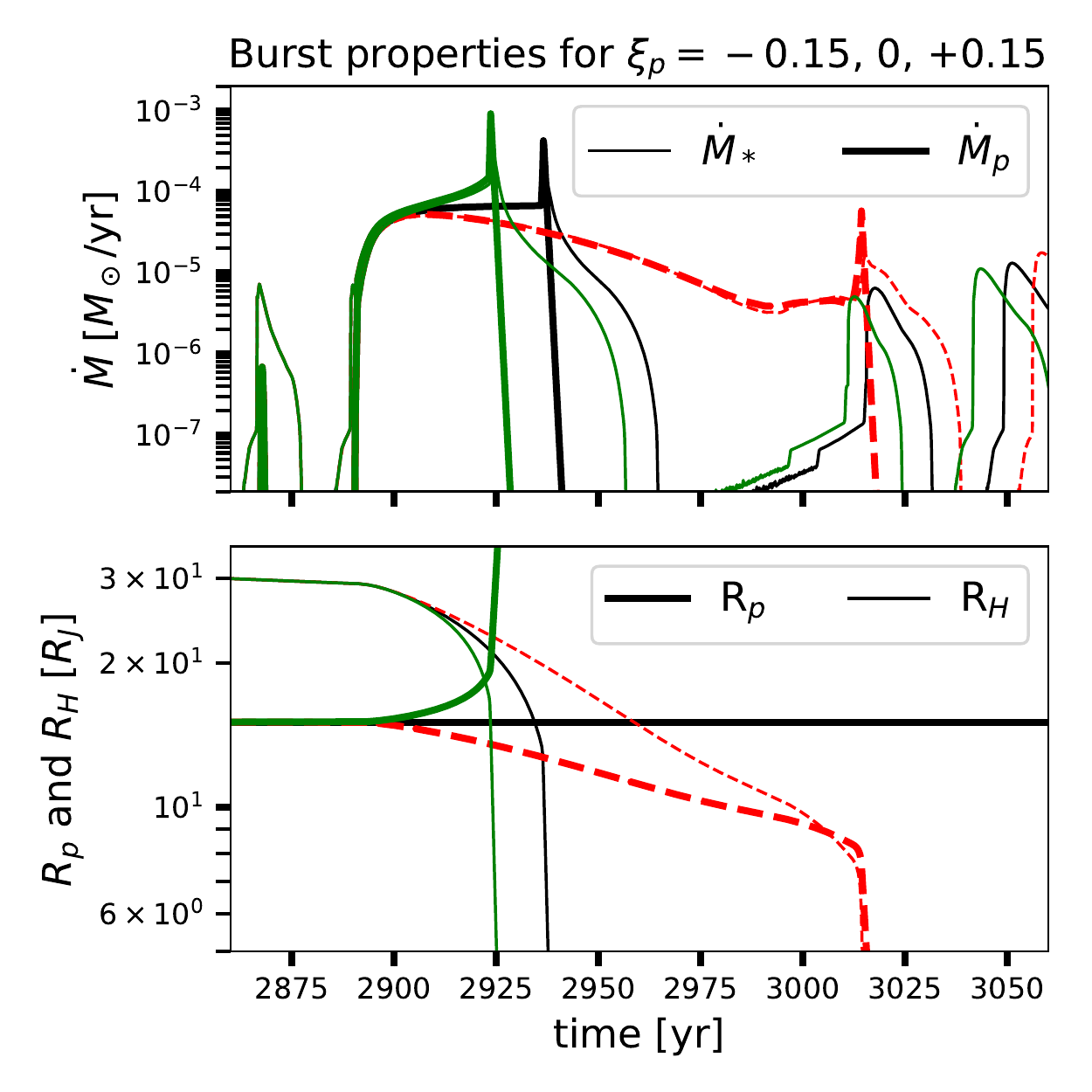}
\caption{Same as the bottom left and the top right panels from Fig. \ref{fig:a0.3_long} but three different power law indices $\xi_{\rm p}$ (see eq. \ref{M-R-relation}). Planets that expand when losing mass ($\xi_{\rm p} < 0$) run out of mass and fill their Roche lobes sooner than planets with $\xi_{\rm p} > 0$. The latter contract as they lose mass.}
\label{fig:a0.3_short}
\end{figure}

Fig. \ref{fig:a0.3_short} zooms in onto $200$~years around the burst from Fig. \ref{fig:a0.3_long}, showing that burst with the red curve in the top panel. 
In addition, we also show two analogous calculations except for $\xi_{\rm p} = -0.15$ and $0$ for the green and black curves, respectively. As the initial radius of the planet is the same for all $\xi_{\rm p}$, FUOR ignition occurs at the same time in the three cases. However at later times $\dot M_{\rm p}$ and $\dot M_*$ evolve differently for different $\xi_{\rm p}$. For $\xi_{\rm p}=0$, the planet radius remains constant, so $\dot M_{\rm p}\approx$~const. For $\xi_{\rm p} = -0.15$ ($+0.15$), the planet expands (contracts), $\dot M_{\rm p}$ increasing (decreasing) in response. For all of the cases shown in Fig. \ref{fig:a0.3_short}, there is a strong but short spike in $\dot M_{\rm p}$ after which it nosedives. This is tidal disruption of the planetary remnant occurring when $R_{\rm H} = a (M_{\rm p}/3 M_*)^{1/3}$ catches up with $R_{\rm p}$. This happens sooner for negative $\xi_{\rm p}$, and thus the tidal disruption burst is more powerful for the green curves in Fig. \ref{fig:a0.3_short}.




\section{A complete model for FU Ori
}\label{sec:complete}


\subsection{Planet origin and properties at $R\sim 0.1$~AU}\label{sec:planet_origin}

FUOR phenomenon is suspected to occur in the earliest $t\lesssim 10^5$ years of protostellar growth \citep[e.g.,][]{AudardEtal14}. This is supported by the fact that discs of FUORs are $3-4$ times more massive than discs in class II sources, with $\sim 2/3$ of them possibly being gravitationally unstable \citep{Kospal21-massive-FUORs}. The only plausible source of massive and extended ($ R_{\rm p} > 10 R_J$) planets at such early times is the disc fragmentation due to gravitational instability (GI) in the outer $R \gtrsim $ tens of AU disc \citep[e.g.,][]{Kuiper51,Boss98}. 

Radiative cooling constraints \citep{Gammie01} show that GI planets are born very far from the inner disc, at $R\gtrsim$ tens of AU \citep{Rafikov05,Rice05}. Analytical estimates \citep{Nayakshin10c} and numerical simulations \citep{VB06,BoleyEtal10,ChaNayakshin11a,ZhuEtal12a,VB15,FletcherEtal19} show that it can take as little as $t_{\rm I}\sim 10^4$ years for GI planets to migrate to the inner $\sim $few AU. This type-I-like planet migration \citep{BaruteauEtal11} terminates when (depending on the planet mass and the value of the $\alpha$ parameter) the planets open deep gaps in the host disc and settle into the type-II regime \citep{FletcherEtal19,HumphriesEtal19}. We note in passing that these pre-collapse planets can also be tidally destroyed if they do not contract rapidly enough \citep[see a review in][]{Nayakshin_Review}, but for FU Ori we are after planets that go through Hydrogen dissociative collapse \citep{Bodenheimer74,GraboskeEtal75}. 

The type II migration timescale \citep[e.g.,][]{SyerClarke95,IvanovEtal99} is
\begin{equation}
    t_{\rm II} \sim \left(1 + \frac{M_{\rm p}}{4\pi \Sigma R^2}\right)  \frac{1}{\alpha \Omega h^2}\;,
    \label{t_mig_2}
\end{equation}
where $\dot M_{\rm feed} = 3\pi \alpha c_s H \Sigma$ is the disc accretion rate behind the planet far from 0.1 AU. The second term inside the brackets in 
eq. \ref{t_mig_2} dominates in the inner disc in our model, and hence 
\begin{equation}
t_{\rm II} \sim \frac{M_{\rm p}}{\dot M_{\rm feed}} \sim 3\times 10^{3} \;\text{years}\; \frac{M_{\rm p}}{3\mj}\;\frac{10^{-6} \msun \text{yr}^{-1}}{\dot M_{\rm feed}}\;
    \label{t_mig_2a}
\end{equation}
At high accretion rates, $t_{\rm II} < t_{\rm I}$, whereas at $\dot M  \lesssim 10^{-7}\msun$~year$^{-1}$ the inverse is true.  GI planets entering the innermost disc regions will be {\em at least} $10^4$~years old.

However, as mentioned previously, we are only interested in post-collapse planets as pre-collapse GI planets cannot make it into the inner 0.1 AU. The ``age" of a post-collapse planet at $\sim 0.1$~AU is then the time it took the planet to migrate from the location at which the pre-collapse molecular clump collapsed, and this has to be beyond an ``exclusion zone"  
\begin{equation}
    R_{\rm exc} \sim \hbox{ 2 AU }\; \left[\frac{M_{\rm p}}{3 \mj} \right]^{2/3}
    \label{R_exclusion}
\end{equation}
\citep[e.g., see the red line in top panel of Fig. 3 in][]{Nayakshin16a}. The contraction age of post-collapse planets is therefore shorter than their age from birth at tens of AU.


What radii will these planets have when they arrive at the inner disc? Dust opacity is a major uncertainty in determining how GI planets contract, both in pre-collapse and post-collapse phases. Concerning post-collapse contraction, in particular, we estimate that dust sedimentation times in the atmospheres of $\sim 10 R_{\rm J}$ gas giant planets are quite short, $\sim 1$~year. Dust-free opacities, usually employed for post-collapse planets, thus appear reasonable. However, pebble accretion on rapidly migrating gas giant planets was found to be very efficient in numerical simulations \citep{BoleyDurisen10,HN18,Forgan19,Baehr19-pebble-accretion}. It can be shown that even a tiny ($\sim 10^{-4}$) fraction of that pebble accretion flux is sufficient to maintain dust opacities in the atmospheres of planets.

\begin{figure}
\includegraphics[width=0.99\columnwidth]{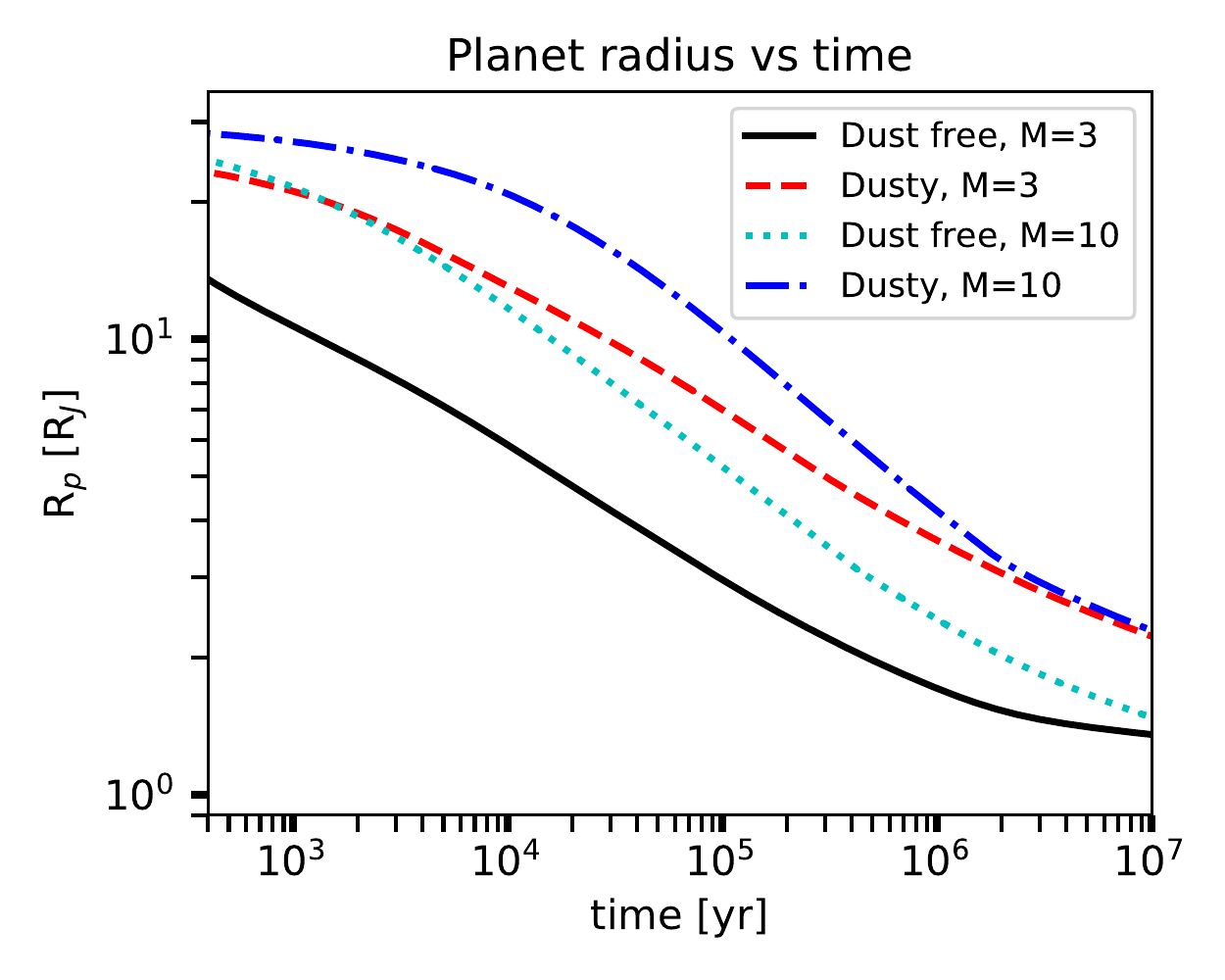}
\caption{Radius vs time for MESA models of post-collapse giant planet contraction. Two values for planet masses are considered, along with either dust-free or dusty opacities. Planets with dusty atmospheres have radii $\sim 10-20 R_{\rm J}$ at ages of a few $10^4$~years.}
\label{fig:MESA_R_vs_t}
\end{figure}

Therefore, we constructed MESA \citep{Paxton13-Mesa-2} models of contracting gas giant planets for both dust-free and dusty atmospheres. For the latter we used dust opacity $\kappa = \kappa_0 (T/1000)^b$ with constants $\kappa_0$ and $b$ varied in a small range consistent with well known opacity calculations \citep[such as][]{SemenovEtal03,Woitke16-DIANA}. In particular, $\kappa_0$ was either 3 or 6, and $b = 0.5$ or 1. We found differences of only $\lesssim 20$\% in planet radius at ages of interest for all these choices, so we present here just the case $\kappa_0=6$ and $b=0.5$. Density-dependent dust sublimation was modelled with the method of \cite{Kuiper10-RHD}. Fig. \ref{fig:MESA_R_vs_t} shows evolution of $R_{\rm p}$ for $M_{\rm p} = 3\mj$ and $10 \mj$ that bracket the plausible range of $M_{\rm p}$. We observe that at the planet age $10^4 \lesssim t \lesssim 10^5 $~years dust-free planets are a factor of 2 smaller than their dusty counterparts. For these ages, dusty planets have radii between $\sim 8 R_{\rm J}$ and $\sim 20R_{\rm J}$, with $10\mj$ planets being larger than $3\mj$ planets by a factor of $1.5$ or so. This implies that on average $10\mj$ planets require roughly twice hotter disc to subject them to extreme evaporation. $10\mj$ planets thus must travel closer to the star to experience EE.

\subsection{Constraints on disc $\dot M_{\rm feed}$}\label{sec:dotM_constraints}

\begin{figure}
\includegraphics[width=0.5\textwidth]{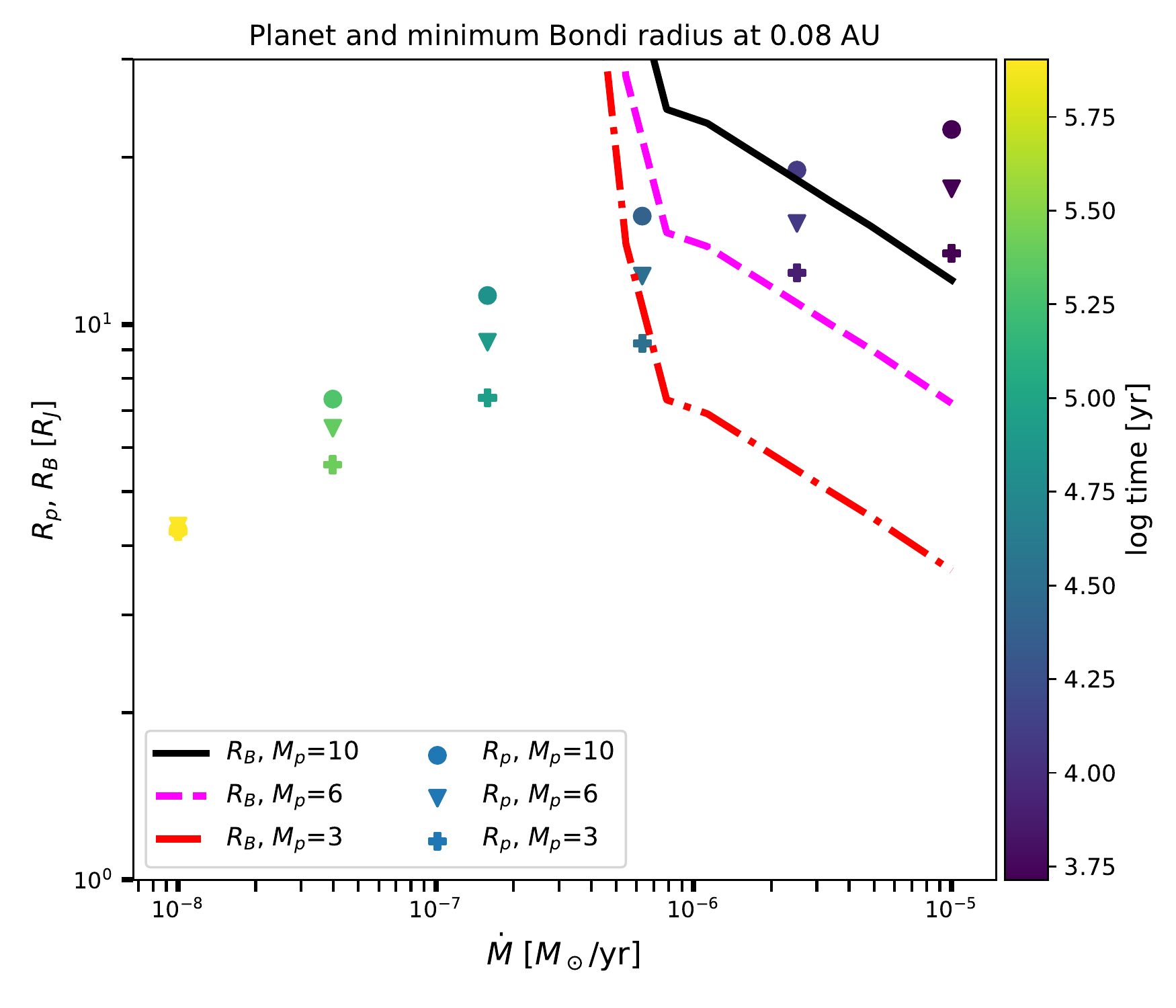}
\caption{Planet radius (symbols) at the time it reaches $R=0.08$~AU, and the minimum Bondi radius (lines) at that location in the disc, versus disc feeding accretion rate. This is plotted for three different values of planet masses. Extreme Evaporation is only possible when $R_{\rm p} > R_{\rm B}$. This favours smaller mass planets migrating at high $\dot M \gtrsim 10^{-6} \msun$~yr$^{-1}$ disc accretion rates.}
\label{fig:Param_space}
\end{figure}

By using the self-consistent planet contraction tracks from \S \ref{sec:planet_origin} we can constraint the disc feeding rate $\dot M$ at large distances. In Fig. \ref{fig:Param_space} we show experiments in which MESA models of planets of masses $M_{\rm p} = 3, 6, 10 \mj$ are injected into the disc at $R=15$~AU. The disc is assumed to be a steady-state \cite{Shakura73} one fed at the rate $\dot M$. The planets then migrate through the disc. Simulations are stopped when the planet reaches $R=0.08$~AU. The radius of the planet {em at that time} is plotted with symbols in Fig. \ref{fig:Param_space} as a function of $\dot M$. Symbol colours convey the time it takes the planet to migrate to $0.08$~AU. For each of the simulations we measure $T_{\rm max}$, the maximum disc temperature at 0.08 AU. At low $\dot M$, the disc is on the lower branch of the S-curve and so $T_{\rm max}\lesssim 3\times 10^3$~K. At high $\dot M$ the disc is unstable to TI, and thus $T_{\rm max}\gtrsim 3\times 10^4$~K. We use $T_{\rm max}$ to define the minimum Bondi radius,
\begin{equation}
    R_{\rm B, min} = \frac{2 G M_{\rm p}\mu}{k_b T_{\rm max}}\;.
    \label{R_B_min}
\end{equation}
$R_{\rm B, min}$ is plotted in Fig. \ref{fig:Param_space} with lines for different $M_{\rm p}$. 

Extreme Evaporation of a planet at $R=0.08$~AU occurs when $R_{\rm p} \ge R_{\rm B, min}$. We see that at low $\dot M$, none of the planets can be affected by EE. The disc does not undergo TI cycles, and also, the planet takes much too long to reach the inner disc (up to 1 Myr), so it contracts too much. Additionally, more massive planets, despite being more radially extended, have a higher $R_{\rm B, min}$ because their gravitational potential is higher. Therefore, $M_{\rm p} = 10\mj$ planet can only be susceptible to EE at $\dot M > 3\times 10^{-6}\msun$~yr$^{-1}$. Moderately massive planets inside a high $\dot M$ discs can go through EE. The high $\dot M$ needed tallies with another requirement for the disc to be massive, i.e., to create the planet via fragmentation.

\subsection{Parameter space: Tidal Disruptions versus Extreme Evaporation}\label{sec:survey}

\begin{figure*}
\includegraphics[width=0.99\textwidth]{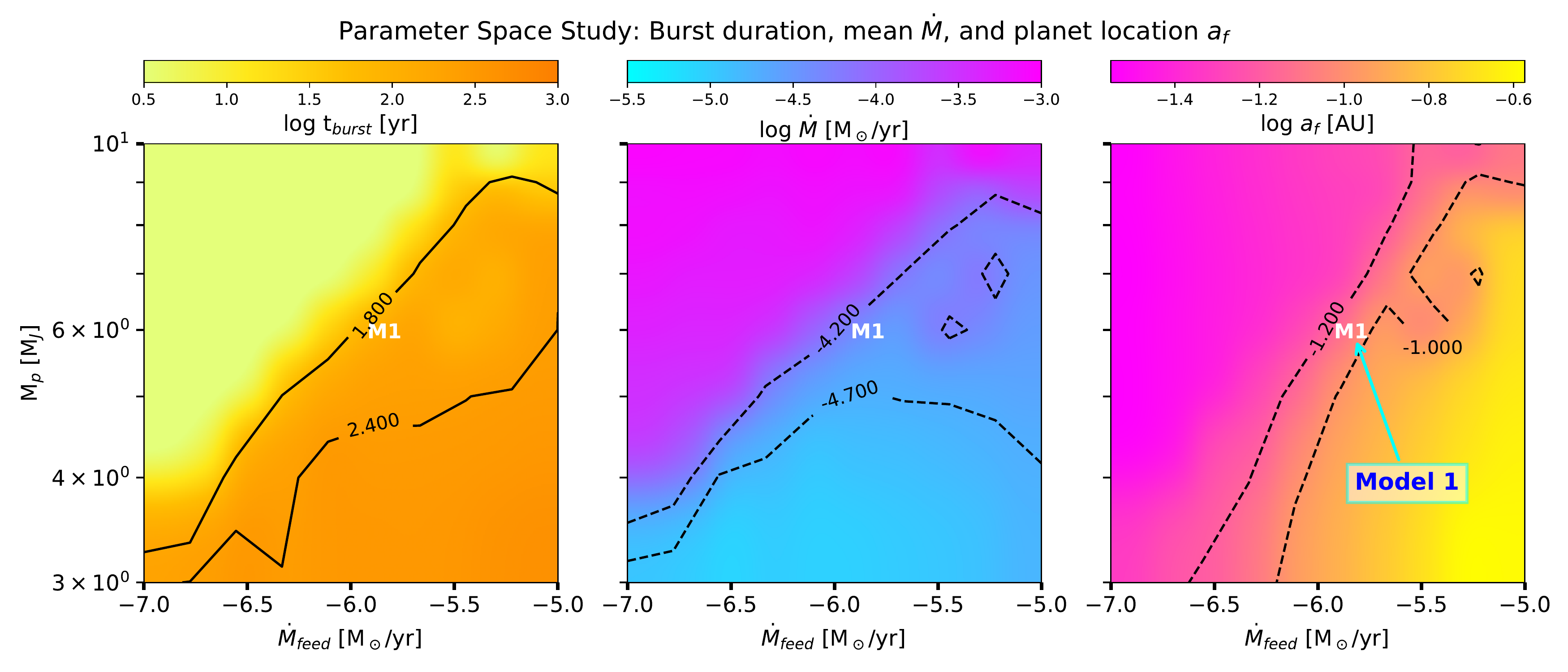}
\caption{Parameter space study (\S \ref{sec:survey}) depicting burst duration (Left panel), mean $\dot M_*$ during the burst (Middle panel), and location where the planet is disrupted/evaporated (Right panel). Symbol ``M1" marks model M1 discussed in \S \ref{sec:model_M1} that appears the best match to the data. The contour lines in the panels mark the approximate region within which good models for FU Ori should be located (see text).}
\label{fig:Param_space_3}
\end{figure*}

We now explore systematically the sensitivity of the model to two major parameters,  the disc feeding rate and the planet's initial mass. We ran a small grid of models in which the former was varied from $\dot M_{\rm feed} = 10^{-7} \msun$~year$^{-1}$ to $\dot M_{\rm feed} = 10^{-5} \msun$~year$^{-1}$, and the latter from $M_{\rm p} = 3\mj$ to $10\mj$. We set $\xi_{\rm p} = 0$, and the starting position for all the planets to $a=10$~AU. Fig. \ref{fig:Param_space_3} presents the most salient results: burst duration (left panel), mean accretion rate during the bursts (middle panel) and the location of planet disruption (right panel). The mean $\dot M$ shown in Fig. \ref{fig:Param_space_3} excludes the tidal disruption spike with which the bursts end (cf. Figs. \ref{fig:a0.3_long}, \ref{fig:a0.3_short}).

The key result from Fig. \ref{fig:Param_space_3} is the existence and a rather clean separation between the two contrasting modes of planet disruption -- Tidal Disruption (TD) and Extreme Evaporation (EE). The two are separated via a diagonal line running from the bottom left corner to the top right one in all of the panels in the figure. 

The top left half of the panels in Fig. \ref{fig:Param_space_3} is filled by most massive planets migrating at all $\dot M_{\rm feed}$, and also by intermediate-mass planets migrating  at low $\dot M_{\rm feed}$ -- these systems tend to suffer TD, which is usually a runaway process \citep[see \S 3 in][]{ElbakyanN-21-HMYSO} with very intense outbursts with $\dot M_* \gtrsim 10^{-3} \msun$~year$^{-1}$ and duration of just a few years.

The bottom right half of the panels in Fig. \ref{fig:Param_space_3} is taken up by ``low mass" (here $M_{\rm p} \leq 4 \mj$) planets, and also by moderately massive giants ($\sim (5-7) \mj$) migrating inside high $\dot M_{\rm feed}$ discs. These planets are exterminated by EE. The resulting accretion bursts last tens to hundreds of years with $\dot M_*$ from a few $\times 10^{-6} \msun$~year$^{-1}$ to $\sim 10^{-4} \msun$~year$^{-1}$.

There is a simple reason for this dichotomy of outcomes. At a similar age, more massive planets are more compact, i.e., their $M_{\rm p}/R_{\rm p}$ ratio is larger. Higher disc temperatures are needed to subject more massive planets to EE, so these planets tend to migrate closer to the star without suffering EE. However, Hill radius shrinks rapidly as planets edge closer to the star, and eventually, TDs occur before EE could occur. In terms of spatial location in the disc, TDs typically occur at $R\sim (0.03 - 0.01) $~AU. Lower-mass planets are more susceptible to EE. As they migrate towards the star, they enter EE regime while $R_{\rm p}$ is still significantly smaller than $R_{\rm H}$. These planets evaporate in a more prolonged steady-state-like fashion, usually at distances $R \sim (0.06 - 0.2)$~AU. We discuss the TD and EE limits in particular examples below.

\subsubsection{Massive planets: short TD bursts}\label{sec:massive_TD}

\begin{figure*}
\includegraphics[width=0.99\textwidth]{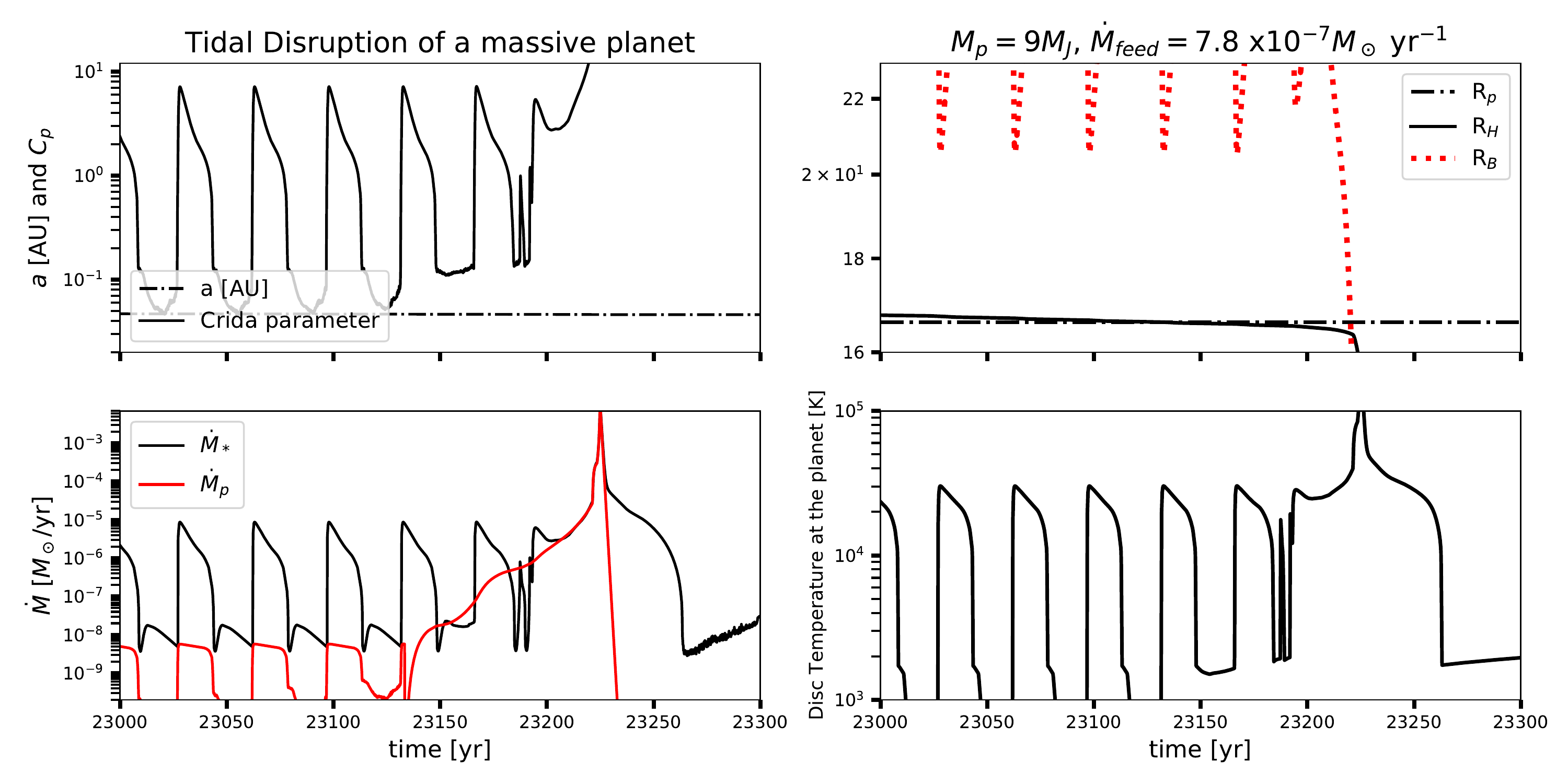}
\caption{An example of a Tidal Disruption (TD) burst due to a massive $M_{\rm p}=9\mj$ planet. The planet is too compact to be susceptible to extreme evaporation ($R_{\rm p} < R_{\rm B}$), but is disrupted tidally by Roche lobe overflow. See \S \ref{sec:massive_TD} for detail.}
\label{fig:JO_a10_massive}
\end{figure*}

Fig. \ref{fig:Param_space_3} shows that most massive planets, $M_{\rm p}\gtrsim 8\mj$ produce short duration (1-10 years) bursts of exceptionally high stellar mass accretion rate, with $\dot M_*$ as high as $10^{-3} \msun$~year$^{-1}$. Here we focus on a specific case of $M_{\rm p} = 9\mj$ planet injected into a disc with $\dot M_{\rm feed} = 7.8 \times 10^{-7} \msun$~year$^{-1}$. Fig. \ref{fig:JO_a10_massive} shows the time close to when the planet is disrupted, which occurs at $a=0.05$~AU. Note that planet mass loss rate is much smaller than $\dot M$ even during the peak TI bursts while $R_{\rm H} \ge R_{\rm p}$ (cf. the top right panel). However, at time $t \sim 23150$~yr the opposite happens, and $\dot M_{\rm p}$ starts to increase extremely rapidly until the planet is destroyed in a fraction of a year \citep[a zoom into the TD burst shows that $\dot M_*$ onto the star is slightly offset and is a little longer due to finite disc viscous time, similarly to the burst in Fig. 2 in][]{ElbakyanN-21-HMYSO}.

\subsubsection{Moderate mass planets: long FU Ori-like bursts}\label{sec:moderate_mass_EE}

\begin{figure*}
\includegraphics[width=0.99\textwidth]{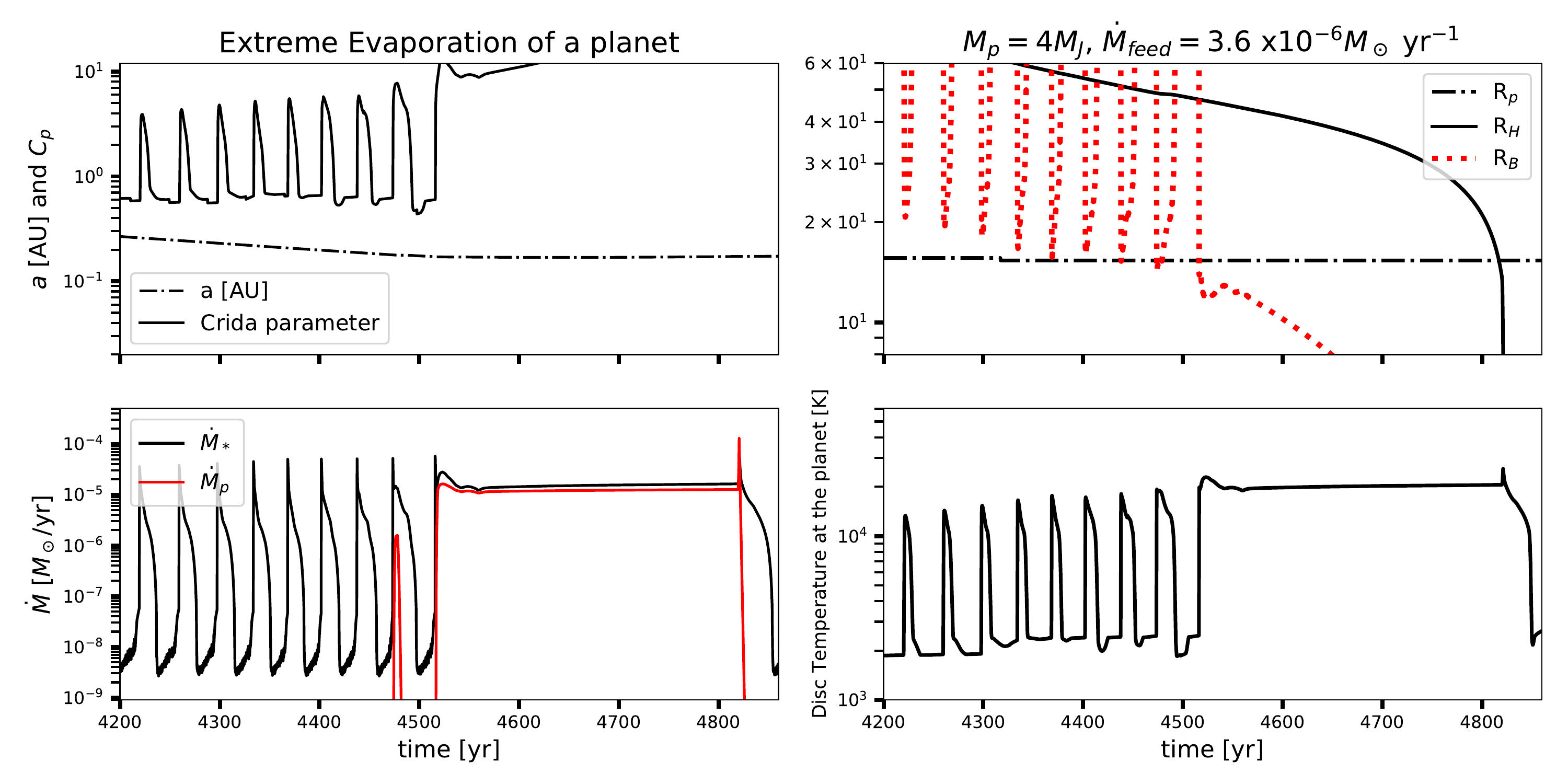}
\caption{Same as Fig. \ref{fig:JO_a10_massive} but for a less massive planet. In this case, the planet enters EE regime at $t\approx 4520$~yr and is eventually destroyed by TD at $t\approx 4830$~yr.}
\label{fig:JO_a10_low_mass}
\end{figure*}

In Fig. \ref{fig:JO_a10_low_mass} we present the case of a $M_{\rm p} = 4\mj$ injected into a disc with $\dot M_{\rm feed} = 3.6 \times 10^{-6} \msun$~year$^{-1}$. The planet becomes prone to the EE process at time $t\approx 4520$~yr when it is at separation $a=0.17$~AU. The top right panel of Fig. \ref{fig:JO_a10_low_mass} shows that at that time $R_{\rm B}$ drops below $R_{\rm p}$ during the TI outburst state. The resulting steady-state EE of the planet lasts for about 300 years until its Hill radius becomes smaller than $R_{\rm p}$. The ``last hurrah" outburst then destroys the planetary remnant in a manner discussed in \S \ref{sec:massive_TD}.

\subsubsection{A standard candle regime in EE bursts}\label{sec:standard_candle}

\begin{figure}
\includegraphics[width=0.49\textwidth]{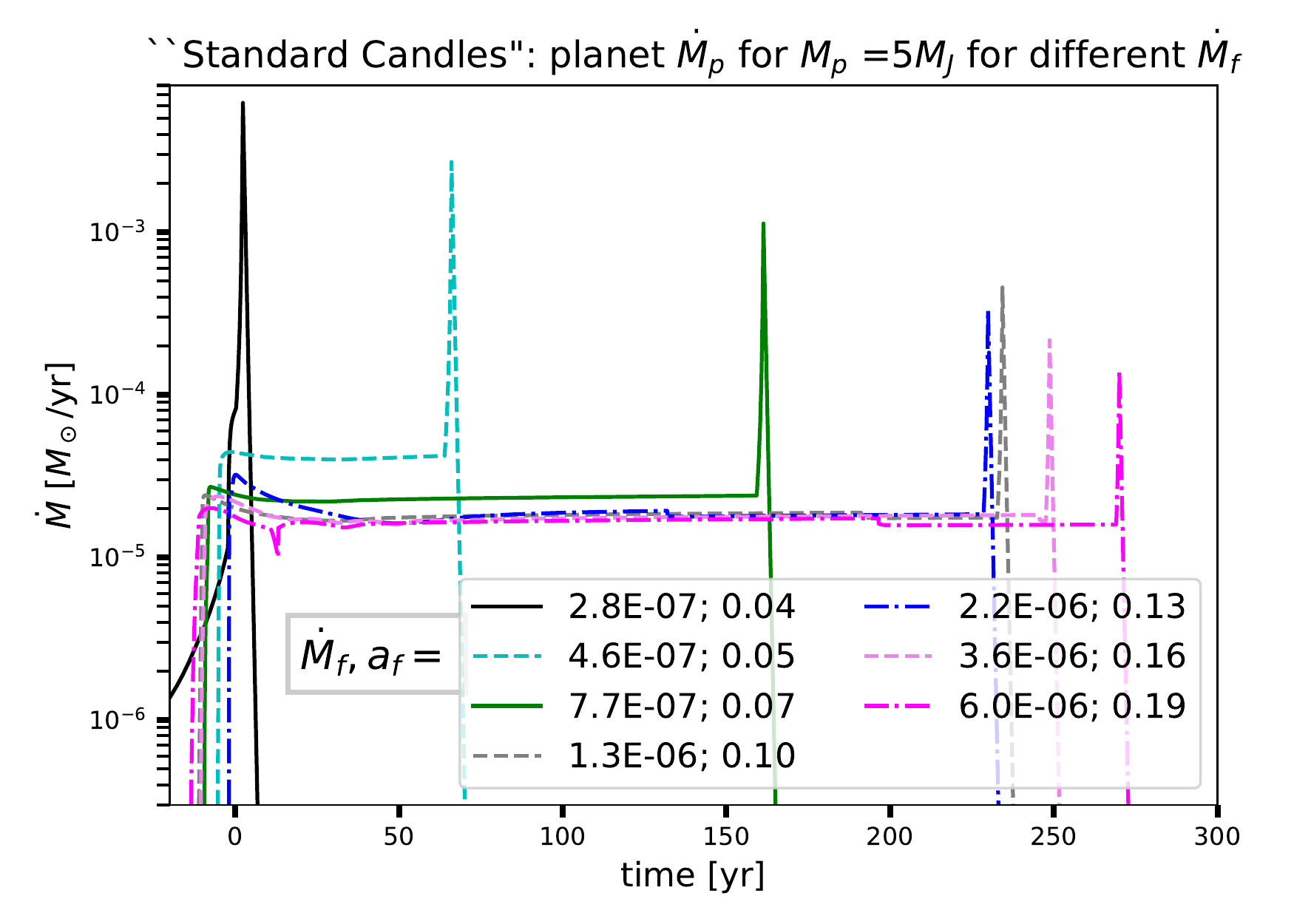}
\caption{Planet mass loss rates for outbursts created by the destruction of planets with mass $M_{\rm p} =5\mj$ migrating through discs with different feeding rates as shown in the legend. These experiments are part of the parameter survey study shown in Fig. \ref{fig:Param_space_3}. As discussed in \S \ref{sec:survey}, planets in discs with low $\dot M_{\rm feed}$ are destroyed via a runaway TD (the black curve), whereas those migrating in younger discs experience a steady-state EE process. Note that $\dot M_{\rm p}$ in the latter case is nearly constant (cf. \S \ref{sec:standard_candle}). The legend gives the disc feeding rate and the location of planet disruption in AU.}
\label{fig:JO_a10_Mp5}
\end{figure}

There is a ``standard candle" regime in EE outbursts that becomes evident when we consider bursts from constant mass planets. Fig. \ref{fig:JO_a10_Mp5} shows the planet mass loss rate for models with $M_{\rm p} = 5\mj$ for a selection of $\dot M_{\rm feed}$. We shifted the time axes so the outbursts start at $t\approx 0$. At the lowest $\dot M_{\rm feed}$, the planet is in the TD regime and is hence disrupted in a few years via run-away Roche Lobe overflow (\S \ref{sec:massive_TD}). 

For all higher values of $\dot M_{\rm feed}$, the burst $\dot M_{\rm p}$ is $\sim (2-4)\times 10^{-5} \msun$~yr$^{-1}$. This is especially so for $\dot M_{\rm feed} > 5\times 10^{-7}\msun$~year$^{-1}$ cases. To understand this, note that eq. \ref{Mdot_fit} gives $\dot M_{\rm p} \propto T_{\rm h}^{\xi} R_{\rm p}^{3/2}$ where $\xi\approx 2.2$. For a general case, both the disc midplane temperature, $T_{\rm h}$, and $R_{\rm p}$ may be expected to vary significantly, so the constancy of $\dot M_{\rm p}$ seen in Fig. \ref{fig:JO_a10_Mp5}, and a weak trend of decreasing $\dot M_{\rm p}$ with increasing $\dot M_{\rm feed}$ may appear paradoxical. However, the EE evaporation sets in when $R_{\rm B}$ drops below $R_{\rm p}$. Planet migrates in slowly {\em in terms of} TI cycles -- generally $\sim 10^2 - 10^3$ TI cycles occur while planet-star separation drops by a factor of 2. Therefore,  $R_{\rm p}$ is very nearly equal to $R_{\rm B}$ when EE begins (cf. the top right panel in Fig. \ref{fig:JO_a10_low_mass}). It then follows that in the self-regulating system that our disc-planet system is, the mass loss rate eq. \ref{Mdot_fit} can be re-written as
\begin{equation}
    \dot M_{\rm candle} = 2\times 10^{-5} 
    \frac{\msun}{\hbox{year}} \left(\frac{M_{\rm p}}{5\mj} \right)^{2.2} \left(\frac{R_{\rm p}}{10 R_{\rm J}}\right)^{-0.7}\;.
    \label{Mdot_standard_candle}
\end{equation}
For a given planet mass the planet's radius varies by about a factor of 2 only when the planet's age varies by a factor of 10 (cf. Fig. \ref{fig:MESA_R_vs_t}). This explains why $\dot M_{\rm p}$ of the bursts in Fig. \ref{fig:JO_a10_Mp5} vary so little for higher $\dot M_{\rm feed}$: their mass loss rate hovers near the ``standard candle" rate given by eq. \ref{Mdot_standard_candle}. This equation also explains the residual trend of decreasing $\dot M_{\rm p}$ with increasing $\dot M_{\rm feed}$ seen in Fig. \ref{fig:JO_a10_Mp5}. At higher $\dot M_{\rm p}$ the planets entering the EE regime are younger (since planet migration time is shorter at higher $\dot M_{\rm feed}$, see \S \ref{sec:planet_origin}), and thus are more extended. Eq. \ref{Mdot_standard_candle} predicts that the more massive the planet is, the shorter the EE outbursts, confirmed by Fig. \ref{fig:Param_space_3}.

\subsubsection{A good fit model M1}\label{sec:model_M1}

\begin{figure*}
\includegraphics[width=0.99\textwidth]{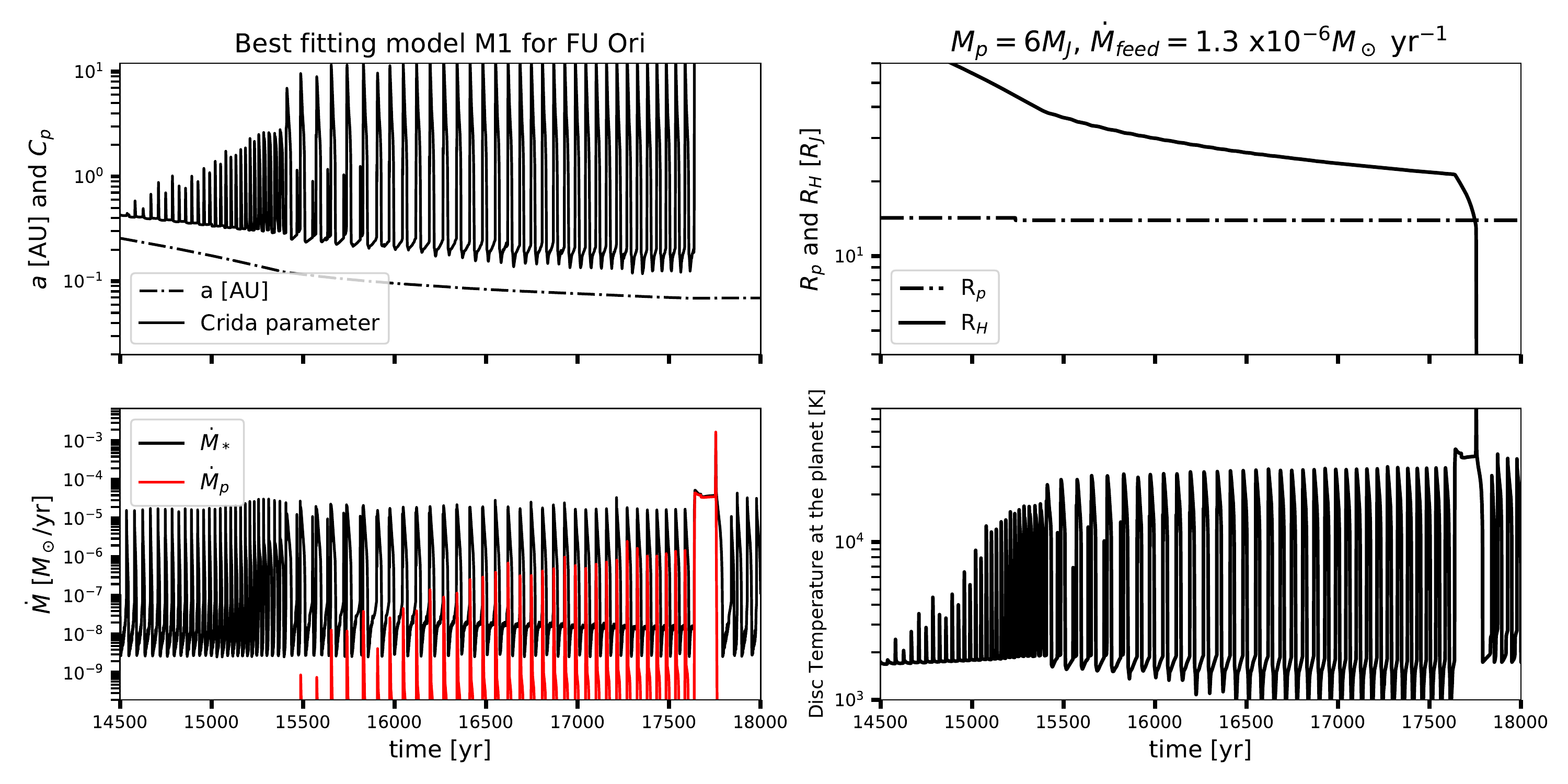}
\caption{A ``good fit" to FU Ori, model M1 marked on Fig. \ref{fig:Param_space_3}.}
\label{fig:JO_a10_best}
\end{figure*}

\begin{figure}
\includegraphics[width=0.49\textwidth]{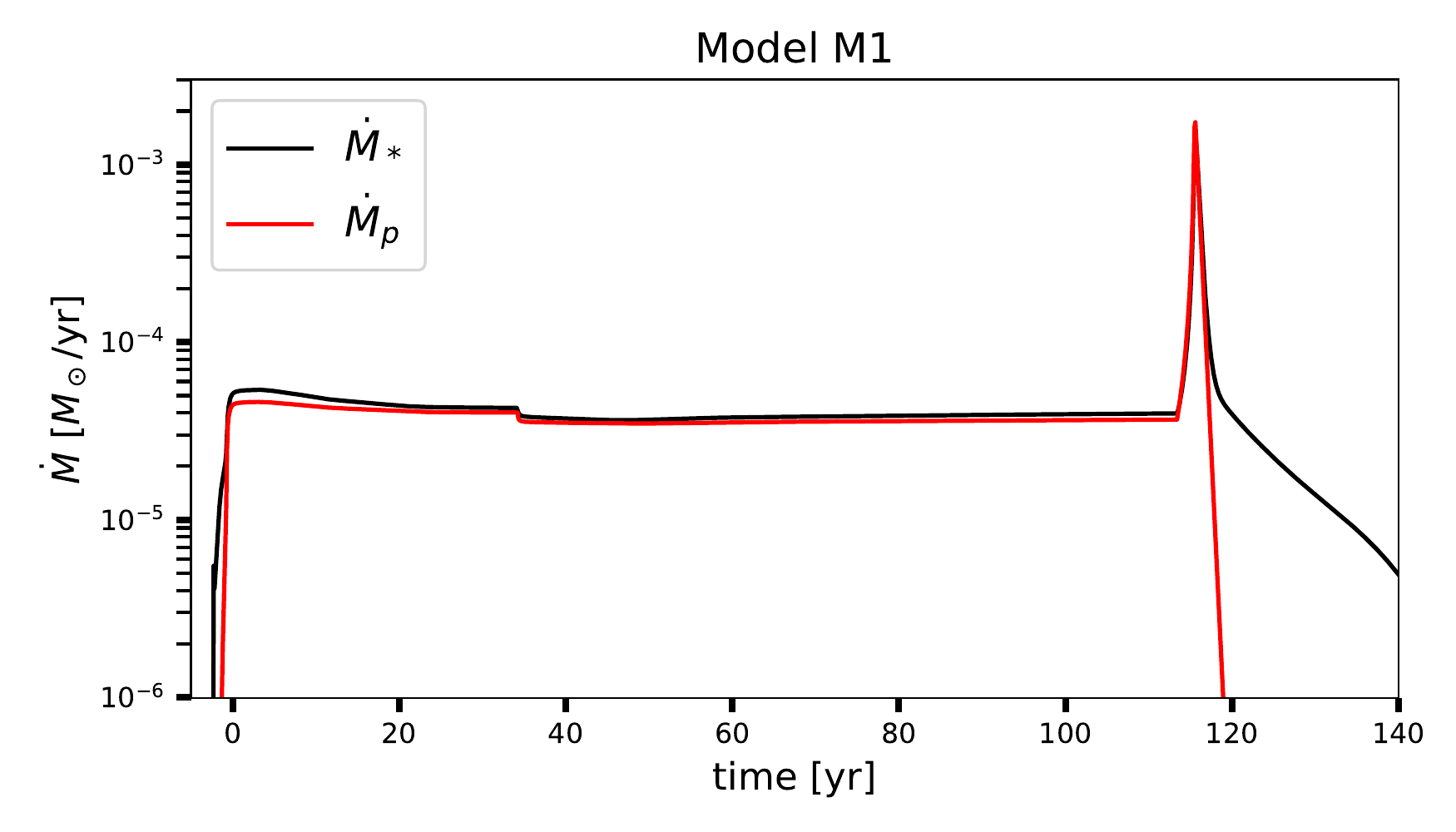}
\caption{Planet mass loss rate and stellar accretion rates vs time for model M1 from fig. \ref{fig:JO_a10_best}.}
\label{fig:JO_Mdot_best}
\end{figure}

Fig. \ref{fig:JO_a10_best} presents a model with $M_{\rm p} = 6\mj$, $\dot M_{\rm feed} = 1.3 \times 10^{-6} \msun$~year$^{-1}$ as the one that comes probably the closest to accounting for the observed properties of FU Ori. EE regime sets in at $a=0.07$~AU where the orbital period is just a little short of what \cite{Siwak21-FUOri-QPOs} deduce. Fig. \ref{fig:JO_Mdot_best} zooms in on the planet $\dot M_{\rm p}$ and star mass accretion rates. The burst has a rise time of a few years, $\dot M_*\sim $ few$\times 10^{-5} \msun$~yr$^{-1}$, declining by about a few tens of \% in its $\sim O(100)$~yrs duration. The outburst ends with a TD of the planet when $M_{\rm p} = 1.4 \mj$.

\subsection{The hot spot}\label{sec:hot_spot}


A hot outflow from a planet is quite likely to result in a hot spot around its location. Midplane temperature in an optically thick disc, $T_{\rm h}$, is typically a few to $\sim 10$ times larger than the disc effective temperature $T_{\rm eff}$. Normally, however, the emission of gas with $T_{\rm h}$ is inaccessible to outside observers as the disc locally shines as a blackbody at $T_{\rm eff}$. However, an outflow from a planet may deliver hot gas to the surface of the disc in planet's vicinity. The process is unlikely to be entirely steady-state which may account for quasi-periodic nature of photometric variability \citep{PowellEtal12,Siwak21-FUOri-QPOs}. 

The luminosity of the perturbation is of order the outflow's mechanical luminosity, $\Delta L \sim \dot M_{\rm p} c_{\rm h}^2$. Compare this with the total disc (accretion) luminosity, $L_{\rm acc} = G M_* \dot M_{\rm p}/R_*$, where we assume that the mass flow rate through the disc $\dot M \sim \dot M_{\rm p}$:
\begin{equation}
	\frac{\Delta L}{L_{\rm acc}} = \frac{\dot M_{\rm p} c_{\rm h}^2 R_*}{G M_* \dot M} \sim 0.01.
	\label{dL}
\end{equation}
Let us estimate the perturbation to the otherwise azimuthally symmetric radiation flux pattern. The unperturbed disc flux at the location of the planet is
$F_0 \sim (3GM_* \dot M_{\rm p})/(8\pi R^3)$.
Due to radiative diffusion, any point-like perturbation to the temperature structure near the disc midplane is smoothed out over an area of the size $A_p \pi H^2$, where $A_p$ is a factor of order a few -- this is the hot spot. The local enhancement of the flux is thus
\begin{equation}
    \frac{\Delta F}{F_0} \sim \frac{\Delta L}{A_p \pi H^2 F_0} = \frac{8}{3 A_p}\sim 1\;.
    \label{Delta_F}
\end{equation}
This is consistent with \cite{Siwak21-FUOri-QPOs} estimate that the effective disc temperature deviates by $\sim 20$\% in the perturbed disc region. In terms of blackbody flux, $\Delta F/F_0 = 4 T_0^3 \Delta T/T_0^4 = 4 \Delta T/T_0 \sim 1$, where $T_0=T_{\rm eff}$, the unperturbed disc effective temperature. 


\section{Discussion}\label{sec:discussion}

We proposed that FU Ori is fed by a massive gas protoplanet located at $a\approx 0.08$~AU from the star and evaporating at an opacity-limited Extreme Evaporation rate, calculated in \S \ref{sec:EE}. Placing a toy fixed radius non-migrating planet in a time-dependent disc (\S \ref{sec:disc-planet}) we showed that EE process is activated during thermal instability (TI) bursts expected to occur in young discs \citep{Bell94} when disc temperatures surrounding the planet exceed $\sim 3\times 10^4$~K. If EE mass loss rate exceeds the local disc accretion rate, then ``planet ignition" occurs. The planet becomes the dominant mass donor for the inner disc, albeit for an astrophysically short period of time.

We then built an approximate but entirely self-consistent planet-in-disc system of equations in \S \ref{sec:complete}, where the planet and disc exchange angular momentum via tidal torques, and energy and mass via EE and/or TD processes. We used MESA to calculate radius evolution for a dust-rich young planet. Moderately massive ($M_{\rm p}$ from a few to $\sim 7\mj$) planets migrating in discs with $\dot M \gtrsim 10^{-6}\msun$~year$^{-1}$ go through a ``standard candle" EE process, producing accretion outbursts with $\dot M_*$ and duration similar to that of classical FUORs. For FU Ori, a small grid of models resulted in a ``good fit model" marked M1 in Fig. \ref{fig:Param_space_3} and discussed in \S \ref{sec:model_M1}. 

Our ideas are related to previous suggestions that TI and/or planets power FUOR events \citep[e.g.,][]{Bell94,LodatoClarke04,Clarke05-FUORs,VB10,NayakshinLodato12}, but require both TI and planets and invoke a previously missed process (EE). Unlike many of prior studies we demand (and succeed) for our disc model to satisfy constraints on disc viscosity derived from  other astronomical systems undergoing TI. We use planet migration and contraction/collapse calculations and/or constraints obtained by other authors in the context not related to FU Ori \citep[e.g., see][for a review]{Nayakshin_Review}. As a result, our model has very few {\em significant} free parameters, e.g., the grid of models in \S \ref{sec:survey} only varies disc feeding rate $\dot M_{\rm feed}$ and $M_{\rm p}$. Yet the model satisfies a number (all that we are aware of at this time) of observational constraints:



\begin{enumerate}
   \item $\dot M$ before and during the burst are $\lesssim 10^{-7}\msun$~year$^{-1}$ and a few $\times 10^{-5}\msun$~year$^{-1}$, respectively (cf. Fig. \ref{fig:JO_a10_low_mass} \& \ref{fig:JO_a10_best}). 
   
    \item Duration of the bursts from tens to hundreds of years. See the middle panel in Fig. \ref{fig:Param_space_3}.
    
    \item The observed hot spot period points at planet location  $a\sim 0.08$~AU. This is naturally explained (not fine-tuned) by our model as the location where TI elicits EE of the planet. The estimated hot spot luminosity is also reasonable (\S \ref{sec:hot_spot}).

    \item The short viscous time at planet location yields outburst rise time of $\sim 1$~year, as observed (Figs. \ref{fig:JO_a10_Mp5}, \ref{fig:JO_Mdot_best}). 
        
    \item The size of the active disc in our model $\lesssim 0.3$~AU. For a planet located at $R\sim 0.1$~AU naturally produces an outer edge of the active disc at $\sim (0.2-0.3)$~AU from the star (Figs. \ref{fig:Disc_FixedMdot_alpha1m2} -- \ref{fig:fluxes_lin}). This is how far the material lost by the planet spreads radially in a burst of $\sim 100$ years duration (\S \ref{sec:Radial_flux_distribution}).
    
    \item Absence of Thermal Ionisation instability during FU Ori observations. The observed light curve of the source has shown a rather steady and mild decline \citep{Clarke05-FUORs}. In \S \ref{sec:no_TI} we showed that this, just like the point (v) above, is naturally understood if there is a quasi-steady source of matter (a planet) in the disc interior to $\sim 0.3$~AU. 
    
    \item Any model for FU Ori should work for $\alpha_{\rm cold} \sim 0.01$ and $\alpha_{\rm hot}\sim 0.1$ (as inferred from other TI-unstable systems, see \S \ref{sec:no_TI}). Our model uses these viscosity choices (\S \ref{sec:complete}).
    
    \item Observations indicate that large amplitude FUOR-like events occur predominantly in massive class 0/I systems \citep{AudardEtal14,Contreras-19-FUOR-statistics}, yet FU Ori's disc is somewhat an outlier in being smaller and less massive than most of FUOR discs \citep{Kospal21-massive-FUORs}. It may thus appear to contradict the requirement for a self-gravitating disc to be present in the system (\S \ref{sec:planet_origin}). However, the physics of our model demands TI in the inner disc, which only occurs at very high $\dot M_{\rm feed} \gtrsim 10^{-6}\msun$~year$^{-1}$. A disc with 10\% of $M_*$ would run out of fuel in less than $\sim 50$ thousand years. This is comparable with the likely planet age. In addition, FU Ori's primary star is more massive and could have photo-evaporated the previously massive self-gravitating disc quite rapidly. Therefore, it is possible that FU Ori disc {\em was} appropriately massive when it fragmented. 
    
    \item  It is estimated that most stars go through a dozen FUOR-like episodes spaced apart by $\sim 10^4$~years \citep{HK96} although see \cite{Fischer-PPVII}. Numerical simulations show from a few to dozens of self-gravitating fragments in massive GI-unstable discs \citep[e.g.,][]{VB06,VB10,ChaNayakshin11a}, which is sufficient to account for observations.
    
\end{enumerate}

In the future, more detailed planet structure models  should be employed. This has a potential to constrain the inner structure of very young planets via FUOR phenomenon. Coupling such models with the disc will allow a wider application of this model to other FUORs, and, potentially, to episodic accretion on high mass stars \citep[e.g.,][]{ElbakyanN-21-HMYSO}.





\section{Aknowledgement}

SN and VE acknowledge the funding from the UK Science and Technologies Facilities Council, grant No. ST/S000453/1. JEO is supported by a Royal Society University Research Fellowship. This project has received funding from the European Research Council (ERC) under the European Union’s Horizon 2020 research and innovation programme (Grant agreement No. 853022, PEVAP). This research used the ALICE High Performance Computing Facility at the University of Leicester, and DiRAC Data Intensive service at Leicester, operated by the University of Leicester IT Services, which forms part of the STFC DiRAC HPC Facility (www.dirac.ac.uk). For the purpose of open access, the authors have applied a Creative Commons Attribution (CC-BY) licence to any Author Accepted Manuscript version arising.

\section{Data availability}

The data obtained in our simulations can be made available on reasonable request to the corresponding author.


\bibliographystyle{mnras}
\bibliography{nayakshin}

\appendix

\bsp	
\label{lastpage}
\end{document}